\documentclass[12pt,a4paper]{article} 
\usepackage{latexsym} 
\usepackage{epsfig} 

\makeatletter 
\@addtoreset{equation}{section} 
\makeatother 
 
\def\unit{\hbox to 3.3pt{\hskip1.3pt \vrule height 7pt width .4pt \hskip.7pt 
\vrule height 7.85pt width .4pt \kern-2.4pt 
\hrulefill \kern-3pt 
\raise 4pt\hbox{\char'40}}}

\def\half{{\textstyle {1 \over 2}}} 
\def\quart{{\textstyle {1 \over 4}}} 
\def\iquart{{\textstyle {i \over 4}}} 
\def\ihalf{{\textstyle {i \over 2}}}

\def\noverm#1#2{{\textstyle {#1 \over #2}}} 
\def\Dpartial{{\cal D}} 
\def\tr{\,{\rm tr}\,} 
 
\def\be{\begin{equation}} 
\def\ee{\end{equation}} 
\def\ba{\begin{eqnarray}} 
\def\ea{\end{eqnarray}} 
 
\newcommand{\beq}{\begin{equation}} 
\newcommand{\eeq}[1]{\label{#1}\end{equation}} 
\newcommand{\ber}{\begin{eqnarray}} 
\newcommand{\eer}[1]{\label{#1}\end{eqnarray}} 
\newcommand{\eqn}[1]{(\ref{#1})}

\def\tr{\,{\rm tr}\,} 
\def\str{\,{\rm str}\,} 
\def\ap{\alpha'} 
\def\at{{\tilde \alpha}'} 
\def\a{\alpha} 
\def\b{\beta} 
\def\g{\gamma} 
\def\G{\Gamma} 
\def\d{\partial} 
\def\e{\epsilon} 
\def\p{\psi}  
\def\c{\chi} 
\def\cb{{\overline\chi}} 
 
\def\m{\mu} 
\def\n{\nu} 
\def\r{\rho} 
\def\l{\lambda} 
\def\lb{{\overline\lambda}} 
\def\k{\kappa} 
\def\s{\sigma} 
\def\o{\omega} 
 
\def\vf{\varphi} 
\def\O{{\cal O}} 
\def\ub{{\overline u}} 
\def\ks{{k \kern-.5em /}} 
\def\es{{\e \kern-.4em /}} 
\def\ds{{\partial \kern-.5em /}} 
\def\Ds{{D \kern-.6em /}}

\def\no{\noindent}

\def \ha {{1\over 2}} 
 
\def\inv{^{\raise.15ex\hbox{${\scriptscriptstyle -}$}\kern-.05em 1}}

\def\makeatletter{\catcode`\@=11}% 11:letter 
\makeatletter 
\def\mathbox#1{\hbox{$\m@th#1$}}% 
\def\math@ccstyles#1#2#3#4#5#6#7{{\leavevmode 
      \setbox0\mathbox{#6#7}% 
      \setbox2\mathbox{#4#5}% 
      \dimen@ #3% 
      \baselineskip\z@\lineskiplimit#1\lineskip\z@ 
      \vbox{\ialign{##\crcr 
             \hfil \kern #2\box2 \hfil\crcr 
             \noalign{\kern\dimen@}% 
             \hfil\box0\hfil\crcr}}}} 
\def\mathaccstyles{\math@ccstyles\maxdimen} 
\def\maththroughstyles{\math@ccstyles{-\maxdimen}} 
\def\unitmatrixDT% 
 {\maththroughstyles{.45\ht0}\z@\displaystyle {\mathchar"006C}\displaystyle 1} 
\begin{document} 
 
\pagestyle{empty} 
\begin{flushright} 
\small 
NEIP-01-003 \\ 
UG/01-29\\ 
VUB/TENA/01/06\\ 
{\bf hep-th/0105274}\\ 
May 2001 
\normalsize 
\end{flushright}

\begin{center} 
 
%title 
 
\vspace{.7cm} 
 
{\Large {\bf Supersymmetric non-abelian Born-Infeld 
 
                 revisited}} 
 
\vspace{.7cm} 
 
\setcounter{footnote}{2} 
%authors 
 E.~A.~Bergshoeff${}^1$, A.~Bilal${}^{2,}$\footnote 
{On leave of absence from Ecole Normale Sup\'erieure,  
24 rue Lhomond 75231 Paris Cedex 05, France} 
, M.~de Roo${}^1$ and A.~Sevrin${}^4$ 
 
\vskip 0.4truecm 
 
 ${}^1$Institute for Theoretical Physics\\ 
   Nijenborgh 4, 9747 AG Groningen\\ 
     The Netherlands\\ 
\vskip 0.4truecm 
 ${}^2$Institute of Physics, University of Neuch\^atel\\ 
   rue Breguet 1, 2000 Neuch\^atel, Switzerland\\ 
\vskip 0.4truecm 
 ${}^4$Theoretische Natuurkunde, Vrije Universiteit Brussel\\ 
   Pleinlaan 2, B-1050 Brussels, Belgium\\ 
 
\vskip 1.0cm 
 
%%%%%%%%%%%%%%%%%%%%%%%%%%%%%%%%%%%%%%%%%%%%%%%%%%%%%%%%%%%%%%%%%%%%%% 
 
{\bf Abstract} 
 
\end{center} 
 
\begin{quotation} 
 
\small 
 
\noindent 
We determine the non-abelian Born-Infeld action, including fermions, 
as it results from the four-point tree-level open superstring 
scattering amplitudes 
at order $\alpha'^2$. We find that, after an appropriate field redefinition 
all terms at this order can be written as a symmetrised trace. 
We confront this action with the results that follow from kappa-symmetry 
and conclude that the recently proposed non-abelian kappa-symmetry 
cannot be extended to cubic orders in the Born-Infeld curvature.

\end{quotation} 
 
\newpage 
 
\pagestyle{plain} 
 
\setcounter{section}{0}

 %title 
 
\section{Introduction} 
 
One of the unsolved questions of D-brane physics concerns the form of the  
(tree-level) effective  
action for $N$ coinciding D-branes beyond the leading term which is just  
${\rm U}(N)$ super Yang-Mills theory.  For a single D-brane, $N=1$, the  
higher-order corrections are captured by (a supersymmetric version of) the  
Born-Infeld Lagrangian \cite{AT, Aga}. Once several D-branes are present, 
things become involved. On the one hand, the gauge field $A_\mu$   
is non-abelian \cite{Witten} and one has to give an ordering  
prescription for the  
higher-order terms. On the other hand, for D$p$-branes, there are also  
$9-p$ embedding coordinates $X^i$ which are ${\rm U}(N)$ valued as well, and  
all background fields will depend on them. This is bound to be quite  
complicated. As a first step, many papers concentrated on  
D9-branes in order to avoid this second difficulty. Of course, the  
D9-brane action is closely related to the open superstring  
effective action with ${\rm U}(N)$ Chan-Paton factors.  
 
The most direct way to obtain the effective action goes through the  
calculation of open string scattering amplitudes. This program yielded 
the purely bosonic terms through order $\alpha'{}^2 F^4$ \cite{GW,Tseyt}.  
The full order $\ap^2$ action in the abelian case was 
determined by \cite{metsaev}. 
It is obvious that 
at higher orders the complexity of this approach considerably 
increases.  Several alternative techniques have  
been developed 
precisely with the aim to avoid these complications.  
 
The most obvious alternative uses $\beta$-function 
calculations. This method proved extremely succesful in the 
abelian case: it was used to show that the (bosonic) Born-Infeld action
is the effective action for the open superstring
theory to all orders \cite{Tseyt}. However, in 
the non-abelian case it becomes as unpleasant as the previous 
approach.  
 
This led to the development of 
several, more indirect ways of attacking the problem.  
Some of them use supersymmetry as a guideline as the supersymmetry 
algebra in 10 dimensions is severely restricted. One obvious choice 
would be to use linear supersymmetry. This was exploited in \cite{BRS},
\cite{goteborg}. In particular, the work of 
\cite{goteborg} led to a full proposal for the effective action through 
order $\ap^2$ including fermionic and derivative terms. The 
presence of a non-linearly realized supersymmetry provided some checks 
on this results and obviously raises the question whether there exists 
an underlying $\k$-invariant action. In the abelian case the answer is 
affirmative. In fact $\k$-symmetry gave the first explicit 
supersymmetrization of the abelian Born-Infeld action in a flat 
background \cite{Aga}. 
 
In \cite{BdRS} the issue of $\k$-symmetry in the non-abelian case was 
addressed. Starting from a concrete ansatz, which was motivated by the abelian 
calculation, this resulted in a $\k$-invariant action including all 
terms quadratic in the field strengths up to quartic fermions.  
 
A perhaps closely related approach uses the existence of BPS-type 
solutions \cite{dks0}. While this method does not give any 
information on the fermionic terms, it does provide a powerful method 
to reconstruct the purely bosonic part of the action. In the abelian 
case it shows that the Born-Infeld action is unique. The extension to 
the non-abelian case is presently under study and will give 
information on the purely bosonic terms through order $\ap^4$ 
including higher-order derivatives \cite{dks0}. 
 
In the context of string theory, configurations involving constant 
magnetic background fields correspond, after T-duality, to D-branes at 
angles. The latter picture allows for a direct calculation of the 
spectrum which can then be compared to the spectrum as calculated from 
the non-abelian Born-Infeld action \cite{wati,dst}.  
Again, this program so far was 
restricted to the study of the bosonic terms only and partially fixed 
the effective action through order $\ap^4F^6$ \cite{stt}.  
 
Finally, the Seiberg-Witten map might give further clues about the 
structure of the higher-order derivative terms \cite{cornalba}. 
 
A major issue in the construction of the effective action in the 
non-abelian case is the ordering of the fields. String theory 
unambiguously determines the $\ap^2F^4$ terms to be a symmetrised 
trace. Modulo effects arising from higher-order derivative terms, this 
led Tseytlin to the conjecture that the full non-abelian Born-Infeld 
action should be defined through the symmetrised trace 
\cite{AT2}. Soon thereafter this proposal was probed by 
comparing fluctuation spectra with those of the corresponding D-brane 
configurations and the result disagreed from order $\ap^4 F^6$ on 
\cite{wati}, \cite{dst}. These results concerned bosonic terms 
only and one might wonder whether fermionic terms at order $\ap^2$ 
already deviate from the symmetrised trace prescription. In \cite{BdRS} 
it was claimed that such a deviation indeed occurs. The claim of 
\cite{BdRS} was based on the assumption that a non-abelian 
generalization of $\k$-symmetry exists. Recent results in 
\cite{goteborg} indicate that the symmetrised trace prescription still 
holds for these fermionic terms at order $\ap^2$.  
 
To settle this issue, we will calculate in Section 2 all terms in the 
effective action, including fermions, which can be determined  
from four-point string scattering amplitudes of 
order $\ap^2$. We 
find that the string effective action, at this order and  
after a certain field redefinition, takes the form 
of a symmetrised trace. Furthermore, it 
agrees with the results in \cite{goteborg}. 
As we will discuss in Section 3, we conclude that the 
non-abelian $\kappa$-symmetry as introduced in \cite{BdRS} does not work 
when cubic orders in the field strength $F$ are included in the variation 
of the action.  
Nevertheless, the presence of a nonlinear supersymmetry in  
\cite{goteborg} suggests the existence of a different formulation, 
perhaps related to $\kappa$-symmetry, in which both  
supersymmetries arise after an appropriate gauge fixing.

\section{Effective action from the string amplitudes} 
 
In this section we will summarize the computation of all the tree-level  
open string (disc) four-point amplitudes between the massless gauge  
bosons and their fermionic partners (gauginos). There is a 4 boson,  
a 4 fermion and a 2 boson / 2 fermion amplitude. We will call the  
external momenta $k_1, \ldots k_4$ (all taken as incoming), assign  
Chan-Paton labels $a,b,c,d= 1, \ldots {\rm dim\ U}(N)$, and  
wave-functions $u_i$ to the external fermions and polarisations  
$\e_j$ to the external bosons. This is depicted in Fig. 1 for the  
example of a 2 boson / 2 fermion amplitude. Our conventions as well as 
various useful identities are summarised in the appendix.

\begin{figure}[h] 
\begin{center} 
\epsfig{figure=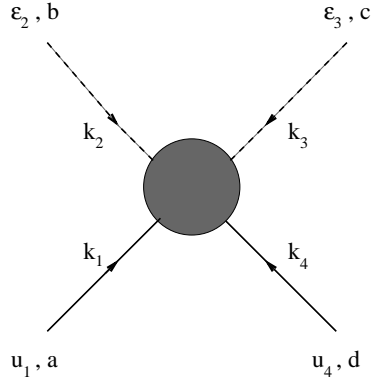} 
\end{center} 
\caption{2 boson / 2 fermion scattering amplitude} 
%\label{fig1} 
\end{figure}

%\begin{figure}[h] 
%\centerline{\epsfig{figure=qft1.eps,  height=6cm, width=6cm}} 
%\caption{2 boson / 2 fermion scattering amplitude} 
%\end{figure} 

\subsection{The string amplitudes} 
 
Any of the 4 point amplitudes is a sum of six disc diagrams  
corresponding to the 6 different cyclic orderings of the vertex  
operators as shown in Fig. 2.

\begin{figure}[h] 
\begin{center} 
\epsfig{figure=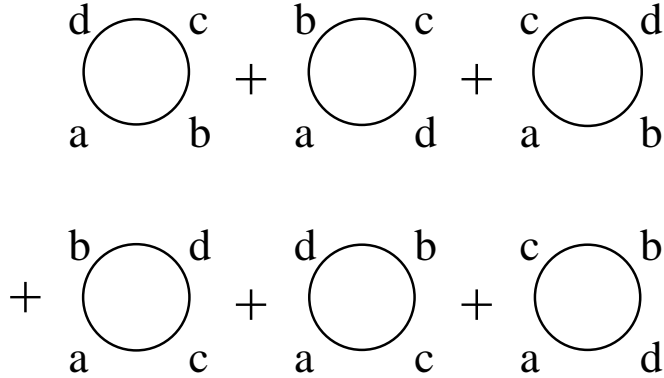} 
\end{center} 
\caption{The six different cyclic orderings} 
%\label{fig2} 
\end{figure}

The contribution of each of the six orderings then is given \cite{GSW,POL}  
by the product of 
 
\no 
1.) a trace of the product of matrices $\l_a$ in the fundamental  
representation of ${\rm U}(N)$, taken in the cyclic order given  
by the diagram of Fig. 2, e.g. for the first one:  
$\tr \l_a\l_b\l_c\l_d \equiv t_{abcd}$ 
 
\no 
2.) a function $G$ depending on the two Mandelstam variables ``flowing"  
through the diagram ``horizontally" and ``vertically". For the first  
diagram of Fig. 2 e.g. the vertical momentum flow gives $(k_1+k_2)^2=s$  
while the horizontal momentum flow gives $(k_1+k_4)^2=u$. Clearly,  
the 1. and 2. diagram give $G(s,u)$, the 3. and 4. give $G(s,t)$ and  
the 5. and 6. give $G(t,u)$. The function $G$ is given by 
\be\label{Gfct} 
G(s,t)=\ap^2 {\G(-\ap s) \G(-\ap t)\over \G(1-\ap s-\ap t)} 
= {1\over st} - {\pi^2 \ap^2\over 6} + \O(\ap^3) 
\ee 
and is the same independent of the nature (boson or fermion) of the  
massless external states. 
 
\no 
3.) a kinematic factor $K$ depending on the polarisations and  
wave-functions in the given cyclic order as well as on the momenta.  
It is independent of $\ap$. This factor would actually be the same also 
for loop amplitudes. In the present example of 2 boson / 2 fermion  
scattering of Fig 1, the 3. diagram of Fig. 2 would e.g. come with a  
$K(u_1, \e_2, u_4, \e_3)$. 
 
\no 
4.) a normalisation factor which we will take to be $-8ig^2$. 
 
\no 
5.) a minus sign for any diagram in Fig. 2 which differs from the first  
one by the permutation of two fermions. Note that these signs will be  
cancelled in the end by the corresponding antisymmetry of the $K$-factor. 
 
Let us now discuss these  kinematical factors $K$. They 
are given in ref. \cite{GSW}, where references to the original
literature can be found.  Some care has to be exercised 
while copying the formula since our conventions are different from those of  
ref. \cite{GSW}. 
The differences are: a) $s_{\rm GSW}=-s$, $t_{\rm GSW}=-u$, $u_{\rm GSW}=-t$, 
b) $\{\G^\m,\G^\n\}_{\rm GSW}= - 2 \eta_{\rm GSW}^{\m\n}$ while we take  
$\{\g^\m,\g^\n\}=  2 \eta^{\m\n}$, and c) we also must change the overall   
normalisation by a factor $-{1\over 4}$ for the 4 fermion and the  
2 boson / 2 fermion case, while in the 4 boson case the GSW normalisation  
is appropriate. 
 
For 4 bosons we get: 
\ba\label{kbos} 
K(\e_1,\e_2,\e_3,\e_4)&=& - {t u\over 4} \e_1\cdot \e_2\, \e_3 \cdot \e_4 
- {s u\over 4} \e_1\cdot \e_3\,  \e_2 \cdot \e_4 
- {s t\over 4} \e_1\cdot \e_4\,  \e_2 \cdot \e_3 \cr 
& & -{s\over 2}{\cal K}_s -{t\over 2}{\cal K}_t  -{u\over 2}{\cal K}_u 
\ea 
where 
\ba\label{ks} 
{\cal K}_s &=& \e_1\cdot k_4\,  \e_3 \cdot k_2\,  \e_2\cdot \e_4 
+ \e_2\cdot k_3\,  \e_4 \cdot k_1\,  \e_1\cdot \e_3  
+ \e_1\cdot k_3\,  \e_4 \cdot k_2\,  \e_2\cdot \e_3 
+ \e_2\cdot k_4\,  \e_3 \cdot k_1\,  \e_1\cdot \e_4\cr 
{\cal K}_t &=& {\cal K}_s \vert_{2\leftrightarrow 3}\cr 
{\cal K}_u &=& {\cal K}_s \vert_{2\leftrightarrow 4} 
\ea 
Note that $K(\e_1,\e_2,\e_3,\e_4)$ is completely symmetric under any  
permutation $i\leftrightarrow j$ and it vanishes if we replace $\e_i$  
by $k_i$ as required by gauge invariance. 
 
For four fermions the $K$-factor is given by 
\be\label{kferm} 
K(u_1,u_2,u_3,u_4)={s\over 8}\, \ub_1\g_\m u_4\,  \ub_2 \g^\m u_3 
- {u\over 8}\,  \ub_1\g_\m u_2\,  \ub_4 \g^\m u_3 \ . 
\ee 
The $u_i$ are the (commuting) ten-dimensional Majorana-Weyl fermion 
wave-functions. Hence we have  
$\ub_i\g^\m u_j = \ub_j \g^\m u_i$ and the Fierz identity 
\be\label{fierz1} 
\ub_1\g_\m u_2\,  \ub_3 \g^\m u_4  
+ \ub_1\g_\m u_3\,  \ub_4 \g^\m u_2  
+ \ub_1\g_\m u_4\,  \ub_2 \g^\m u_3  = 0 
\ee 
which together with the relation $s+t+u=0$ implies that  
$K(u_1,u_2,u_3,u_4)$ is completely antisymmetric under the exchange of any  
two fermions, e.g. we have $ K(u_1,u_2,u_4,u_3)=-K(u_1,u_2,u_3,u_4)$ etc. 
 
For two fermions and two bosons, ref. \cite{GSW} considers two cases  
separately:  
the two fermions are adjacent or not. Both cases actually lead to the  
{\it same} $K$-factor: 
\be\label{kferbos} 
K(u_1,\e_2,\e_3,u_4)=K(u_1,\e_2,u_4,\e_3)= 
{u\over 8}  A + {s\over 8} B 
\ee 
where we define the convenient expressions ($\ks \equiv k_\m \g^\m$) 
\ba\label{AB} 
A&=& \ub_1 \es_2 (\ks_3+\ks_4)\es_3 u_4 \cr 
B&=& 2 \ub_1 \left( \es_3 \, k_3\cdot\e_2 -\es_2 \, k_2\cdot \e_3  
- \ks_3\, \e_2\cdot \e_3 \right) u_4 \ .  
\ea 
Using the on-shell properties 
$k_2\cdot \e_2 = k_3\cdot\e_3 = \ks_4 u_4 = \ub_1 \ks_1 =0$ 
one easily shows 
\ba\label{ABsym} 
A\vert_{2\leftrightarrow 3} &=& A-B \quad , \quad  
A\vert_{1\leftrightarrow 4} = B-A \cr 
B\vert_{2\leftrightarrow 3} &=& -B \quad\quad , \quad  
B\vert_{1\leftrightarrow 4} = B \ , 
\ea 
so that they are symmetric under the  
exchange of the two bosons and antisymmetric under exchange of the  
two fermions. 
 
These kinematical factors are actually determined by the required  
(anti)symmetry, (linearized) gauge invariance and dimensional considerations. 
 
It follows that any of the four-point (tree-level) amplitudes we are  
interested in takes the form 
\ba\label{A4gen} 
A_4&=&-8ig^2\ K(1,2,3,4) \ \times\cr 
&\times&\left\{ 
(t_{abcd}+t_{dcba}) G(s,u) 
+(t_{abdc}+t_{cdba})G(s,t) 
+(t_{acbd}+t_{dbca})G(t,u)  \right\} \ .\cr 
& & 
\ea 
Note that any minus signs introduced when two fermions in Fig. 2 are  
permuted with respect to the reference configuration has been cancelled  
by another minus sign when performing the same permutation on the   
arguments of $K$ to rewrite it as $K(1,2,3,4)$. 
 
Turning to the traces, they come in 3 combinations: 
\ba\label{indivtraces} 
T_1&=& t_{abcd}+t_{dcba}= 
{1\over 2} \left(d_{abe}d_{cde}+d_{ade}d_{bce}-d_{ace}d_{bde}\right) \cr 
T_2&=&t_{abdc}+t_{cdba}= 
{1\over 2} \left(d_{abe}d_{cde}+d_{ace}d_{bde}-d_{ade}d_{bce}\right) \cr 
T_3&=&t_{acbd}+t_{dbca}= 
{1\over 2} \left(d_{ace}d_{bde}+d_{ade}d_{bce}-d_{abe}d_{cde}\right) \ . 
\ea 
where $d_{abc}$ is given by $\{\l_a, \l_b\}=d_{abc} \l_c$.  
Properties of the $d$ and $f$ tensors are given 
in the appendix. Note that the symmetrised trace is given by 
\ba\label{tracecomb} 
\str \l_a\l_b\l_c\l_d  
= {1\over 12}  
\left( d_{abe}d_{cde} + d_{ace}d_{bde} + d_{ade}d_{bce}\right)\ . 
\ea 
 
Inserting the $\ap$-expansion of the $G$-function into \eqn{A4gen} we get 
for any of the four-point amplitudes  
\be\label{A4exp} 
A_4=-8ig^2\ K(1,2,3,4) \sum_{n=0}^\infty a_4^{(n)} \ap^n\ . 
\ee 
The lowest order term can be written in 3 equivalent ways: 
\be\label{a40} 
a_4^{(0)} 
= {1\over s} \left( {1\over t} f_{ace}f_{bde}  
+ {1\over u} f_{ade}f_{bce}\right)  
= -{1\over u} \left( {1\over s} f_{abe}f_{cde}  
+ {1\over t} f_{ace}f_{bde}\right)  
= {1\over t} \left( {1\over s} f_{abe}f_{cde}  
- {1\over u} f_{ade}f_{bce}\right) \ . 
\ee 
This vanishes in the abelian case: there is no lowest order photon-photon  
scattering. Clearly, there is no order $\ap$ contribution and $a_4^{(1)}=0$. 
 
The obvious fact about the order $\ap^2$ contribution is that it is always  
a symmetrised trace. Indeed, at order $\ap^2$ the function $G$ is just a  
constant, and thus all traces contribute equally, leading to a symmetrised  
trace: 
\be\label{a42} 
a_4^{(2)}= -\pi^2 \str\l_a\l_b\l_c\l_d \ . 
\ee 
Clearly, there is no reason for any other $a_4^{(n)}$ to be a  
symmetrised trace. 
Note that nevertheless, by construction,  
all $a_4^{(n)}$ are completely symmetric under exchange of any two  
external states, so that the symmetry properties of the amplitude are  
correctly given by those of the kinematical factors $K(1,2,3,4)$. 
 
For convenience of comparison with the field theory amplitudes, we  
explicitly write down the amplitudes up to and including the order  
$\ap^2$ terms: 
 
\underline{four bosons} 
\ba\label{A44b} 
A_4^{\rm 4b}  
&=& \Bigg[ 4ig^2\, {1\over s} \left( {\cal K}_t-{\cal K}_u  
+{u-t\over 4} \e_1\cdot\e_2\, \e_3\cdot\e_4 \right) f_{abe}f_{cde} \cr 
& &  -ig^2\, \e_1\cdot\e_2\, \e_3\cdot\e_4  
\left(  f_{ace}f_{bde} +  f_{ade}f_{bce}   \right)\Bigg]     
+ [ 2 \leftrightarrow 3, b  \leftrightarrow  c]  
+ [ 2 \leftrightarrow 4, b  \leftrightarrow  d] \cr 
&+&  8ig^2 \pi^2 \ap^2 K(\e_1,\e_2,\e_3,\e_4) \str \l_a\l_b\l_c\l_d  
+\O(\ap^3) \ . 
\ea

\underline{two bosons and two fermions} 
\ba\label{A42b2f} 
A_4^{\rm 2b/2f}  
&=& -ig^2\, {B\over u}\,  f_{ade}f_{bce}  + ig^2\, {A\over s}\, f_{abe}f_{cde} 
+ ig^2\, {A-B\over t}\, f_{ace}f_{bde} \cr 
&+& ig^2 \pi^2 \ap^2\, (u A + s B) \str \l_a\l_b\l_c\l_d + \O(\ap^3)  
\ea 
 
\underline{four fermions} (using the Fierz identity) 
\ba\label{A44f} 
A_4^{\rm 4f}  
&=& \Big[ -ig^2\, \ub_1\g^\m u_2\, \ub_4 \g_\m u_3\, {1\over s} f_{abe}f_{cde}  
+  ig^2\, \ub_1\g^\m u_3\, \ub_4 \g_\m u_2\, {1\over t} f_{ace}f_{bde} \cr 
& &  -ig^2\, \ub_1\g^\m u_4\, \ub_2 \g_\m u_3\, {1\over u} f_{ade}f_{bce}  
\Big] \cr 
&+&  ig^2 \pi^2 \ap^2 \left( s\, \ub_1\g^\m u_4\, \ub_2 \g_\m u_3 
- u \, \ub_1\g^\m u_2\, \ub_4 \g_\m u_3  \right) \str \l_a\l_b\l_c\l_d  
+ \O(\ap^3)  
\ea 
 
%%%%%%%%%%%%%%%%%%%%%%%%%%%%%%%%%%%%%%%%%%%%%%%%%%%%%%%%%%%%%%%%%%%%%%%%%% 
 
\subsection{The ansatz for the effective action} 
 
\subsubsection{The $\ap$ expansion} 
 
Our goal is to find the effective action which reproduces the $\ap$-expansion  
of the open superstring four-point amplitude of the previous subsection. At  
lowest order in $\ap$ this is of course well-known to be the ${\rm U}(N)$  
${\cal N}=1$ super Yang-Mills theory in ten dimensions 
\be\label{syn} 
{\cal L}_{\rm SYM} 
=\tr \left( -{1\over 4} F_{\m\n}F^{\m\n} +{i\over 2} \cb \g^\m D_\m \c\right). 
\ee 
This fixes the normalisations of the fields. 
The main effort in this section will be devoted to (almost)  
uniquely determining the effective action  
at order $\ap^2$. Of course the action will be determined only up to terms  
that vanish ``on shell". Since we look at four-point amplitudes we in 
principle fix all terms of the form $F^4$, $F^2\c^2$ and 
$\c^4$ including all higher order derivatives. 
 
The possible terms at order $\ap^2$ are given by dimensional analysis:  
in any space-time  
dimension,  dimensionless quantities are  
$\ap g F_{\m\n}$ and $\ap^2 g^2\, \cb \g D \c$ where $g$ is the 
Yang-Mills coupling constant.

\subsubsection{The abelian Born-Infeld action} 
 
For the sake of comparison we now give the  
expansion of the {\it abelian} Born-Infeld action \cite{Aga}: 
\ba\label{bi} 
{\cal L}_{\rm BI} &=& {1\over \at^2} \left\{ 1-\left[ \det \left( 
\eta^\m_{\ \n} +\at F^\m_{\ \n} - i \at^2\, \cb \g^\m \d_\n\c 
-{\at^4\over 4}\, \cb \g^\r \d^\m\c\, \cb \g_\r \d_\n\c \right)  
\right]^{1/2} \right\} \cr 
&=& -{1\over 4} F_{\m\n}F^{\m\n} +{i\over 2} \cb \ds\c  
+ {i\over 2} \at\, \cb \g_\m \d_\n\c F^{\m\n} \cr 
&& +{\at^2\over 8} \left( F_{\m\n}F^{\n\r}F_{\r\s}F^{\s\m}  
-{1\over 4} \left( F_{\m\n}F^{\m\n}\right)^2 \right)  
 + {i\over 2} \at^2\, \cb \g_\m \d_\n\c  
\left( F^{\m\r}F_{\r}^{\ \n} - \eta^{\m\n} F_{\r\s}F^{\r\s} \right) \cr 
&&+{\at^2\over 4} \left( {1\over 2} \cb\g^\m\d^\n\c\, \cb \g_\m\d_\n\c 
- \cb\g^\m\d^\n\c\, \cb \g_\n\d_\m\c  
+{1\over 2} \left( \cb\ds\c\right)^2 \right) 
+\O(\at^3)  
\ea 
where  
\be\label{alphatilde} 
\at=2\pi g \ap \ . 
\ee 
Note that our present computation of four-point amplitudes at order 
$\ap^2$ will not be sensitive to terms of the form 
$\at^2 (\cb\ds\c)^2$ and $\at^2 \cb\ds\c F_{\r\s}F^{\r\s}$ as they  
vanish on-shell. Nevertheless, they could be determined from 
higher-point amplitudes. Henceforth we will drop such terms. 
A similar  remark applies to the order  
$\at$ term $\cb \g_\m \d_\n\c F^{\m\n} $ which upon partial integration  
and using the Majorana properties can be written as  
$A^\m \cb (\g_\m \d^2 -\d_\m \ds)\c$ which vanishes for on-shell fermions.  
Thus this term does not contribute to a three-point amplitude, but it gives  
a non-vanishing contribution to the 2 fermion / 2 boson four-point amplitude  
via a one-particle reducible diagram with an internal fermion line. 
 
This order $\ap$ term in the abelian Born-Infeld action can be removed by  
the field redefinition 
\be\label{abelfieldredef} 
\c\to \c+{1\over 4} \at F_{\r\s} \g^{\r\s} \c 
\ee 
at the expense of modifying the order $\ap^2$ terms. Dropping all terms  
involving $\at^2 \ds\c$ we then get 
\ba\label{biafter} 
{\cal L'}_{\rm BI}\vert_{\rm on-shell}  
&=& -{1\over 4} F_{\m\n}F^{\m\n} +{i\over 2} \cb \ds\c  
+{\at^2\over 8} \left( F_{\m\n}F^{\n\r}F_{\r\s}F^{\s\m}  
-{1\over 4} \left( F_{\m\n}F^{\m\n}\right)^2 \right) \cr 
&& + {i\over 4} \at^2\, \cb \g_\m \d_\n\c  F^{\m\r}F_{\r}^{\ \n}  
-{i\over 8} \at^2 \cb \g_{\m\n\r} \d_\s \c F^{\m\n}F^{\r\s}\cr 
&&+{\at^2\over 8} \cb\g^\m\d^\n\c\, \cb \g_\m\d_\n\c 
- {\at^2\over 4} \cb\g^\m\d^\n\c\, \cb \g_\n\d_\m\c  
+\O(\at^3)  
\ea 
Note the modified coefficient of the $\cb\g\d\c FF$-term and the new term  
involving $\g_{\m\n\r}$. 
 
There are two four-fermion terms, but they are related by a Fierz  
transformation: 
\be\label{quarticfermions} 
 \cb\g^\m\d^\n\c\ \cb \g_\n\d_\m\c  
\simeq  {2\over 3}\ \cb\g^\m\d^\n\c\ \cb \g_\m\d_\n\c  
\ee 
where $\simeq$ means equality up to on-shell terms. 
This is most easily seen to follow from \eqn{fierztwo} by setting 
$\p=\d_\n\c$, $\l=\d_\m\c$ and $\vf=\c$: dropping on-shell terms 
and using also \eqn{fierzone} 
this becomes 
\ba\label{abelianfierz} 
\cb\g^\m \d_\n\c \d_\m\cb \g^\n\c  
&\simeq& 
{1\over 8}  \cb\g^{\r\s\m}\c\ \d_\m\cb \g_{\n\r\s} \d^\n\c 
-{1\over 48} \cb\g^{\r\s\l}\c\ \d_\m\cb \g_{\r\s\l} \d^\m\c\cr 
&\simeq& 
- \cb\g^\m\d^\n\c\ \cb\g_\m\d_\n\c  
+ \ha \cb\g^\m\d^\n\c\ \cb\g_\n\d_\m\c 
\ea 
from which follows \eqn{quarticfermions}. 
 
\subsubsection{The ansatz for the non-abelian effective action} 
 
We write the effective action as 
\be\label{effaction} 
{\cal L}={\cal L}_{\rm SYM} + {\cal L}_{\rm 4b}+ {\cal L}_{\rm 2b/2f} 
+ {\cal L}_{\rm 4f} + {\cal L}_* + \O(\ap^3 g^2, \ap^2 g^3) 
\ee 
with ${\cal L}_{\rm 4b}$, ${\cal L}_{\rm 2b/2f}$ and ${\cal L}_{\rm 4f}$ 
containing the order $\ap$ and $\ap^2$ terms needed to reproduce the string  
amplitudes to this order. The piece ${\cal L}_*$ contains any terms   
$\sim \at^2 \Ds\c$, $\sim \at^2 D_\m \cb \g^\m$, $\sim \at^2 D_\m F^{\m\n}$  
that vanish on-shell and do not contribute to the four-point amplitudes  
as discussed above. In the following we write ${\cal L}_1 \simeq {\cal L}_2$  
if  ${\cal L}_1$ and ${\cal L}_2$ only differ up to terms in ${\cal L}_*$  
and up to partial integration. 
The meaning of $\O(\ap^3 g^2, \ap^2 g^3)$ is the following: an $n$-point 
(tree) amplitude comes with a factor $g^{n-2}$ and neglecting 
$\O(\ap^2 g^3)$ terms is tantamount to not taking into account terms 
that only contribute to five- and higher-point amplitudes. On the 
other hand, $\O(\ap^3 g^2)$ terms arise from four-point amplitudes 
but contain more derivatives and will not be considered here either.
 
The \underline{purely bosonic piece} ${\cal L}_{\rm 4b}$ is well-established: 
\be\label{fourbosl} 
{\cal L}_{\rm 4b} = \at^2 \str \left(  
{1\over 8} F_{\m\n}F^{\n\r}F_{\r\s}F^{\s\m}  
-{1\over 32} \left( F_{\m\n}F^{\m\n}\right)^2 \right)  
\ee 
where $\at=2 \pi g \ap$. Since this contains exactly four $F$'s, the  
contribution to the four gluon amplitude is obtained by extracting the  
interaction where each $F^a_{\m\n}$ is replaced simply by  
$\d_\m A_\n^a-\d_\n A_\m^a$. There is then a single order $\ap^2$ four  
gluon vertex contributing to the amplitude, and it is a straightforward  
exercise to show that the result coincides with the order $\ap^2$ part  
of the string amplitude $A_4^{\rm 4b}$ in \eqn{A44b}. 
In fact, it is not necessary to check all the terms in  
$K(\e_1,\e_2,\e_3,\e_4)$ since the structure of $K$ is fixed by gauge  
invariance and permutation symmetry. It is e.g. enough to check that  
\eqn{fourbosl} yields the  
$\e_1\cdot\e_2\, \e_3\cdot\e_4$ term  with the correct coefficient. 
It is also easy to show that eq. \eqn{fourbosl} with the symmetrised  
trace is the {\it unique} interaction that reproduces the string amplitude  
at this order. 
 
More interesting is the \underline{mixed piece}  
${\cal L}_{\rm 2b/2f}$. Taking into account the  
Majorana-Weyl properties (see appendix), we find that a general  
ansatz for the non-abelian effective action at orders $\ap$ and $\ap^2$ is 
\ba\label{bosfermlagr} 
{\cal L}_{\rm 2b/2f}&=& i c_1 \at d_{abc} \cb^a \g_\m D_\n \c^b F^{c\m\n} 
+i \at^2 \o_{abcd} \cb^a \g_\m D_\n \c^b F^{c\m\r}F^{d\ \n}_{\ \r}\cr 
&&\cr 
&&+ i \at^2 \xi_{abcd} \cb^a\g_{\m\n\r}\left( D_\s \c^b F^{c\m\n}F^{d\r\s} 
- D^\r\c^bF^{c\m\s} F^{d\ \n}_{\ \s} \right) \cr 
&&\cr 
&\simeq& i c_1 \at d_{abc} \cb^a \g_\m D_\n \c^b F^{c\m\n} 
+i \at^2 y_{abcd} \cb^a \g_\m D_\n \c^b F^{c\m\r}F^{d\ \n}_{\ \r}\cr 
&&\cr 
&&+ i \at^2 \xi_{abcd} \cb^a\g_\m\g_\n\g_\r  D_\s \c^b F^{c\m\n}F^{d\r\s} 
\ea 
with 
\be\label{yabcd} 
y_{abcd}=\o_{abcd}-\xi_{abcd}-\xi_{abdc} \ . 
\ee 
We have not specified the gauge structure of the order $\at^2$ terms:  
$\o_{abcd}$ and $\xi_{abcd}$ are arbitrary so far\footnote{The
ansatz  (\ref{bosfermlagr}) in its present form still contains
terms that vanish on-shell, and which really belong in 
${\cal L}_*$. We will come back to this point after $y$ and
$\xi$ have been matched to the string amplitude results.}. 
On the other hand, 
for the order $\at$ term we have specified $d_{abc}=\str\l_a\l_b\l_c$.  
The only other possibility would be $f_{abc}$. In this latter case however,  
$f_{abc}\cb^a \g_\m D_\n \c^b F^{c\m\n} 
\simeq -{1\over 2}f_{abc} \cb^a \g_\m  \c^b D_\n F^{c\m\n}$. 
% Such a term is possible in the non-abelian case, as is also  
% $f_{abc} \cb^a \g_{\m\n\r} D^\r \c^b F^{c\m\n}$ (while the analogous 
% $d_{abc} \cb^a \g_{\m\n\r} D^\r \c^b F^{c\m\n}$ vanishes upon partial  
% integration by virtue of the Bianchi identity). While these order  
% $\at$ terms $\sim f_{abc}$ are a priori possible they cannot be simply  
% eliminated by a field redefinition of the type discussed below in eq. 
% \eqn{nonabelfieldredef} and hence are expected to give a non-vanishing  
% contribution to the four-point amplitude at order $\ap$ which is not  
% present in the string amplitude.  
In order to somewhat simplify our  
discussion we will assume from the outset that they are not present  
and we start with an action as given by \eqn{bosfermlagr}. 
 
As in the abelian case, the order $\at$ term does not contribute to a  
three-point amplitude between on-shell states, which is consistent with  
the absence of such an amplitude in string theory. It is  
convenient to first eliminate this order $\ap$  
term by performing a field redefinitions and then compute the amplitude. So  
we let 
\be\label{nonabelfieldredef} 
\c^a\to\c^a+{c_1\over 2}\at\, d_{abc} F^b_{\r\s}\g^{\r\s} \c^c\ . 
\ee 
This will not change ${\cal L}_{\rm 4b}$ or ${\cal L}_{\rm 4f}$  but it will 
affect ${\cal L}_{\rm 2b/2f}$ which becomes (up to on-shell terms and  
total derivatives) 
\ba\label{bosfermlagrter} 
{\cal L'}_{\rm 2b/2f}&=& 
i \at^2 y_{abcd} \cb^a \g_\m D_\n \c^b F^{c\m\r}F^{d\ \ \n}_{\ \r} 
+ i \at^2 z_{abcd} \cb^a\g_\m\g_\n\g_\r  D_\s \c^b F^{c\m\n}F^{d\r\s}\cr 
&&\cr
&=& 
i\at^2{\widetilde y}_{abcd} \cb^a \g_\m D_\n \c^b F^{c\m\r}F^{d\ \ \n}_{\ \r} 
+ i \at^2 z_{abcd} \cb^a\g_{\m\n\r}  D_\s \c^b F^{c\m\n}F^{d\r\s} 
\ea 
with 
\ba\label{zabcd} 
z_{abcd}&=&\xi_{abcd}-{c_1^2\over 2} d_{ace}d_{bde} \ , \cr 
{\widetilde y}_{abcd}&=&y_{abcd}+ 2 z_{abcd}  
= \o_{abcd} +\xi_{abcd}-\xi_{abdc}-c_1^2 d_{ace}d_{bde}\ . 
\ea 
 
It is clear from these relations  
that a symmetrised trace presciption can hold at best for  
${\cal L}_{\rm 2b/2f}$ {\it or} 
${\cal L'}_{\rm 2b/2f}$, but not both. 
Note that any part of $y_{abcd}$ or ${\widetilde y}_{abcd}$
that is antisymmetric in $a$ and $b$ and symmetric in $c$ and $d$, 
vanishes on shell by virtue of the Bianchi identity for $F$. 
Hence we can assume $y_{[ab](cd)}={\widetilde y}_{[ab](cd)}=0$
 
For the \underline{four fermion interaction} ${\cal L}_{\rm 4f}$ we take  
the ansatz 
\be\label{lfourferm} 
{\cal L}_{\rm 4f} = \at^2 g_{abcd} \cb^a \g^\m D^\n \c^b\, \cb^c \g_\m D_\n \c^d 
+ \at^2 h_{abcd} \cb^a \g^\m D^\n \c^b\, \cb^c \g_\n D_\m \c^d \ . 
\ee 
Other terms could be written down e.g. 
$ j_{abcd} \cb^a \g^{\m\n\r} D_\r \c^b\, \cb^c \g_\m D_\n \c^d$ 
or 
$ l_{abcd} \cb^a \g^{\m\n\r} D^\s \c^b\, \cb^c \g_{\m\n\r} D_\s \c^d$. 
However, using Fierz identities, all of them can be rewritten as  
\eqn{lfourferm}, up to on-shell terms. 
Similarly, one may assume that $h_{abcd}$ is symmetric under  
interchange of $a$ and $b$, or of $c$ and $d$, or of $ab$ and $cd$, and that 
\be\label{gsymprop} 
g_{abcd}=g_{cdab} \quad {\rm and} \quad 
g_{(ab)[cd]}=g_{[ab](cd)}=0 \quad \Rightarrow \quad g_{abcd}=g_{badc} \ . 
\ee 
 
Finally note that the only order $\ap$ term would be a new fermion bilinear  
like $\at \cb ^a \g^{\m\n\r} D_\m D_\n D_\r \c^a$. It would give an order  
$\ap$ two fermion - one gluon vertex and would contribute to the four  
fermion scattering via one-particle reducible gluon exchange diagrams,  
but this term actually reduces to the term discussed below (\ref{yabcd}).

\subsection{Matching the amplitudes} 
 
The four gluon amplitude has already been discussed above. 
 
\subsubsection{Matching the 2 boson / 2 fermion amplitude} 
 
The most convenient form of the  
relevant interaction is the first line of 
\eqn{bosfermlagrter}, i.e. \underline{after}  
the field redefinition \eqn{nonabelfieldredef}. Indeed, \eqn{bosfermlagrter}  
only contributes two terms to the 2 boson / 2 fermion interaction,  
obtained upon replacing  
$D_\l\to \d_\l$ and $F^a_{\m\n}\to \d_\m A^a_\n - \d_\n A^a_\m$.  
Obviously, there is no order $\a'$ piece, while the computation of  
the order $\ap^2$ contribution to the amplitude is a  bit lengthy  
but straightforward. We get: 
\ba\label{qftbosfermampl} 
A_4^{\rm 2b/2f}&=& i \at^2 \Bigg\{ A( t z^++s z^-)\cr 
& & \phantom{i\at^2} + \ub_1 \es_3 u_4  
\left[ 2 k_1\cdot \e_2 ( t z^+ +s z^-)  
+ {1\over 2} (t k_1\cdot \e_2 - s k_4\cdot \e_2) (y_{dacb}+y_{adcb}) \right] \cr 
& & \phantom{i\at^2} - \ub_1 \es_2 u_4  
\left[ 2 k_4\cdot \e_3 ( t z^+ +s z^-)  
+ {1\over 2} (t k_4\cdot \e_3 - s k_1\cdot \e_3) (y_{dabc}+y_{adbc}) \right] \cr 
& & \phantom{i\at^2} - \ub_1 \ks_3 u_4  \e_2\cdot \e_3 
\left[ -2 t z^+ +{s\over 2} (y_{adbc}+y_{dacb})  
- {t\over 2} (y_{dabc}+y_{adcb})\right] \cr 
& & \phantom{i\at^2} + \ub_1 \ks_3 u_4  
\Big[ k_1\cdot \e_2 k_1 \cdot \e_3 (y_{dabc}-y_{dacb}) 
+  k_4\cdot \e_2 k_4 \cdot \e_3 (y_{adcb}-y_{adbc}) \cr 
& & \phantom{i\at^2+ \ub_1 \ks_3 u_4} 
- k_1\cdot \e_2 k_4 \cdot \e_3 (4z^- +y_{adbc}+y_{dacb})\cr 
& & \phantom{i\at^2+ \ub_1 \ks_3 u_4} 
+ k_4\cdot \e_2 k_1 \cdot \e_3 (4z^+ +y_{dabc}+y_{adcb}) \Big] \Bigg\} 
\ea 
with 
\be\label{zplus} 
z^+=z_{dabc}+z_{adcb} \quad , \quad z^-=z_{adbc}+z_{dacb} 
\ee 
and where $A$ (and $B$) where defined in \eqn{AB}. As a first check, note  
that this indeed vanishes if we replace $\e_i\to k_i$, as required by gauge  
invariance. 
 
As a further consistency check, note that we could have started with the  
interaction 
\eqn{bosfermlagr}, i.e. \underline{before} the field redefinition  
\eqn{nonabelfieldredef}. Then the term $\sim y_{abcd}$ contributes as  
above in \eqn{qftbosfermampl} while the contribution of the term  
$\sim\xi_{abcd}$ can also be read from \eqn{qftbosfermampl} by  
replacing $z_{abcd}\to \xi_{abcd}$ and analogously  
$z^+\to \xi^+=\xi_{dabc}+\xi_{adcb}$ and $z^-\to\xi^-=\xi_{adbc}+\xi_{dacb}$.  
But now we have in addition the contributions from the order $\at$ piece.  
This yields a new order $\at$  cubic (2 fermion - 1 gluon) and a new  
quartic (2 fermion - 2 gluon) vertex. Thus we get various non-vanishing  
contributions to the 2 boson / 2 fermion amplitude at order $\at$, but  
they sum up to zero 
\be\label{orderapampl} 
A_4^{\rm 2b,2f}\vert_{{\rm order}\ \at} =0 
\ee 
in agreement with the string amplitude. 
At order $\at^2$ there are the two diagrams with an internal fermion in  
the $s$ or $t$ channel and both vertices being the cubic order $\at$  
interaction. Their sum yields 
\ba\label{asplusat} 
A_s^{\at^2} + A_t^{\at^2}  
&=& - i c_1^2 \at^2  \Bigg\{ \left(A + 2 k_1\cdot \e_2  \ub_1 \es_3 u_4 
- 2 k_4\cdot \e_3 \ub_1 \es_2 u_4  \right) 
 ( t d_{ace}d_{bde} +s d_{abe}d_{cde}) \cr 
& & \phantom{i\at^2} + \ub_1 \ks_3 u_4  \e_2\cdot \e_3 
2 t d_{ace}d_{bde}\cr 
& & \phantom{i\at^2} + \ub_1 \ks_3 u_4  
\left(4 k_4\cdot e_2 k_1 \cdot \e_3  d_{ace}d_{bde} 
- 4  k_1\cdot e_2 k_4 \cdot \e_3 d_{abe}d_{cde} \right) \Bigg\} \ . 
\ea 
This has to be added to the contributions $\sim y_{abcd}$ and  
$\sim \xi_{abcd}$ as obtained from \eqn{qftbosfermampl} as discussed  
above. Not too surprising, we find that the result of adding this  
contribution is just to shift 
\be\label{shift} 
\xi^+\to\xi^+ -c_1^2 d_{ace}d_{bde} \quad , \quad  
\xi^-\to\xi^- -c_1^2 d_{abe}d_{cde} 
\ee 
without affecting the $y_{abcd}$. This corresponds to 
\be\label{shifttwo} 
\xi_{abcd}\to \xi_{abcd} - {c_1^2\over 2} d_{ace}d_{bde} 
\ee 
which is nothing but replacing $\xi_{abcd}$ by $z_{abcd}$. Thus in the  
end we get exactly the same result \eqn{qftbosfermampl} as obtained from  
the interaction \eqn{bosfermlagrter} \underline{after} the field redefinition. 
 
Matching the result \eqn{qftbosfermampl} to the corresponding string  
amplitude \eqn{A42b2f} (recall that $\at=2\pi g \ap$) 
yields the following conditions 
\be\label{matching} 
z^+=z^- = -{1\over 4} \str \l_a\l_b\l_c\l_d \quad , \quad 
y_{adbc}+y_{dabc} =  \str \l_a\l_b\l_c\l_d \quad , \quad 
y_{adbc}= y_{adcb} \ . 
\ee 
This can be equivalently written as 
\be\label{matchingbis} 
z_{abcd}+z_{badc}= -{1\over 4} \str \l_a\l_b\l_c\l_d \quad , \quad  
y_{(ab)cd} =  {1\over 2} \str \l_a\l_b\l_c\l_d  \quad , \quad  
y_{ab[cd]}=0 \ . 
\ee 
Using the results of the appendix on the general form of 4-index tensors  
arising from a single trace, the most general solution is 
(recall $y_{[ab](cd)}=0$)
\ba\label{matchsol} 
y_{abcd} &=& {1\over 2}  \str \l_a\l_b\l_c\l_d + y_5 d_{cde}f_{abe}\cr 
z_{abcd} &=& -{1\over 8} \str \l_a\l_b\l_c\l_d  + z_4\, d_{abe}f_{cde}  
+ z_5\, d_{cde}f_{abe}  
\ea 
where $y_5$, $z_4$ and $z_5$ are undetermined parameters. 
 
The presence of the undetermined parameters $y_5$, $z_4$ and $z_5$ is
related to the fact that the tensors $y$ and $z$ were not a priori
restricted to avoid the presence of contributions in the ansatz
(\ref{bosfermlagr}) that vanish on-shell. In fact, all three unknown
parameters give contributions to the effective action that vanish  
on-shell, and can be eliminated by field redefinitions.

In the contribution $y_5$, one uses
\begin{eqnarray}
  &&  y_5 d_{cde}f_{abe}\cb^a \g_\m D_\n \c^b F^{c\m\r}F^{d\ \ \n}_{\ \r}
  = \half D_\n(\cb^a\g_\m\c^b) F^{c\m\r}F^{d\ \ \n}_{\ \r}
\nonumber\\
  &&\quad \simeq -\half \cb^a\g_\m\c^b (D_\n F^{c\m\r})F^{d\ \ \n}_{\ \r}
   =  +\quart \cb^a\g_\m\c^b (D^\m F^{c\r}{}_\n)F^{d\ \ \n}_{\ \r}
   \simeq 0\,,
\end{eqnarray}
where the last step requires a partial integration. The contribution
$z_5$ contains the same term as $y_5$, and in addition
\begin{eqnarray}
   z_5 d_{cde}f_{abe} \cb^a\g_{\m\n\r} D_\s\c^b F^{c\m\n}F^{d\r\s}
  \simeq -\quart z_5 d_{cde}f_{abe} 
      \cb^a\g_{\m\n\r\s\l} D^\l\c^b F^{c\m\n}F^{d\r\s}\,,
\end{eqnarray}
where we have used the fact that the product of the two $F$'s is, due
to the symmetry in $cd$, completely antisymmetric in $\m\n\r\s$. The
expression then vanishes due to the Bianchi identity for $F$ after partial 
integration.
The contribution $z_4$ requires a cancellation between the two 
contributions in (\ref{bosfermlagrter}). The trick here is to write
\be
   2d_{abe}f_{cde}\cb^a\g_\m D_\n\c^b F^{c\m\r}F^{d\ \ \n}_{\ \r}
  \simeq - d_{abe}f_{cde} \cb^a\g_{\m\n\l} D^\l\c^b F^{c\m\r}F^{d\ \ \n}_{\ \r}
\ee
and then to do a partial integration in both terms in
(\ref{bosfermlagrter}). The cancellation occurs because
\be
  f_{cde}( D^{[\l} (F^{c\m\s}F^{d\ \ \n]}_{\ \s}) - 
      D_\s (F^{c[\m\n}F^{d\l]\s}) )
  \simeq 0
\ee
where antisymmetrization is over the indices $\m\n\l$.

The matching of the four-point amplitude has therefore 
completely determined the 2 boson / 2 fermion part of the effective action.
 
\subsubsection{Matching the 4 fermion amplitude} 
 
There are again two possible types of contributions to the four fermion 
amplitude: one-particle irreducible diagrams coming from the quartic  
interactions of ${\cal L}_{\rm 4f}$, eq. \eqn{lfourferm},   and, possibly,  
one-particle reducible  
gluon exchange diagrams using the cubic vertex from the order $\at$ 
term in ${\cal L}_{\rm 2b/2f}$ before the field redefinition. This cubic  
vertex however vanishes if both fermions are on shell, so that these gluon  
exchange diagrams do not contribute to the 4 fermion amplitude. This is  
consistent with the fact that the field redefinition does not affect  
${\cal L}_{\rm 4f}$. In particular also, there is no order $\ap$  
contribution to the amplitude. 
 
We will now discuss the contributions of the two terms in ${\cal L}_{\rm 4f}$  
to the  
4 fermion amplitude. We will argue soon that the second term in  
${\cal L}_{\rm 4f}$ cannot reproduce anything that looks like the  
string amplitude unless it can be transformed - using some Fierz  
identity - into a term with the same Lorentz index structure as   
the first one in  
${\cal L}_{\rm 4f}$. So we begin by examining the contribution of this  
first term alone.  Obviously, its contribution to  
the amplitude contains $ \ub_1 \g_\m u_2 \ub_3 \g^\m u_4$, 
$ \ub_1 \g_\m u_4 \ub_2 \g^\m u_3$ and  
$ \ub_1 \g_\m u_3 \ub_2 \g^\m u_4$. Using the Fierz identity 
\eqn{fierz1} this last expression can be rewritten as a combination  
of the two other, and, upon taking into account \eqn{gsymprop} we get 
\ba\label{a44fg} 
A_4^{\rm 4f}\vert_{g-{\rm terms}} 
&=&-2 i\at^2 \Bigg\{  
\left[ (g_{acbd}+g_{adbc})\, s-g_{adcb}\, t-g_{acdb}\, u\right] 
\ub_1 \g_\m u_4 \ub_2 \g^\m u_3 \cr 
& & \phantom{-2 i\at^2 \Bigg\{   } 
-\left[ (g_{acdb}+g_{abdc})\, u-g_{abcd}\, t-g_{acbd}\, s\right] 
\ub_1 \g_\m u_2 \ub_3 \g^\m u_4 \Bigg\} \ . 
\ea 
Also in this case we will not attempt to restrict $g$ a priori to
avoid terms that vanish on-shell.
Comparing with the string amplitude we find that, if and only if 
\be\label{gcond} 
g_{abcd}=g_{acbd} \ , 
\ee 
the amplitude reduces to the desired form 
\be\label{a44fgbis} 
A_4^{\rm 4f}\vert_{g-{\rm terms}} 
=-2 i\at^2 \left( g_{acbd}+g_{abdc}+g_{adcb}\right) 
\left( s\ \ub_1 \g_\m u_4 \ub_2 \g^\m u_3  
- u\ \ub_1 \g_\m u_2 \ub_3 \g^\m u_4 \right) \ . 
\ee 
The symmetry requirements \eqn{gcond} and \eqn{gsymprop} on $g_{abcd}$  
 and the results of  
the appendix on 4-index tensors determine it to be of the form 
\be\label{gform} 
g_{abcd}={g_1\over 12} \left(d_{abe}d_{cde}+d_{ace}d_{bde}\right)  
+ {g_3\over 12}d_{ade}d_{bce}  
\ee 
which in turn implies that  the contribution of the first term in  
${\cal L}_{\rm 4f}$ to the 
amplitude can  be written as 
\be\label{a44fgter} 
A_4^{\rm 4f}\vert_{g-{\rm terms}} 
=-2 i\at^2 (2 g_1 + g_3)  \str \l_a\l_b\l_c\l_d  
\left( s\ \ub_1 \g_\m u_4 \ub_2 \g^\m u_3  
- u\ \ub_1 \g_\m u_2 \ub_3 \g^\m u_4 \right) \ . 
\ee 
 
Next, we consider the second term in  ${\cal L}_{\rm 4f}$. As discussed  
above, we may assume $h_{abcd}=h_{bacd}=h_{abdc}$. A straightforward  
computation shows that it contributes terms like  
$\ub_1 \ks_3 u_2\, \ub_3 \ks_1 u_4$ to the amplitude which are not of the  
desired form  $ \ub_1 \g_\m u_2 \ub_3 \g^\m u_4$ etc.  
However, we will now show that if also $h_{dbca}=h_{abcd}$  
then by a Fierz transformation these terms actually have the desired form.  
Note that requiring $h_{dbca}=h_{abcd}$ together with   
$h_{abcd}=h_{bacd}=h_{abdc}$ implies that $h_{abcd}$ is completely symmetric  
in all its indices, i.e. it is proportional to $\str\l_a\l_b\l_c\l_d$. 
Clearly, once we assume that $h_{abcd}\sim \str\l_a\l_b\l_c\l_d$, the  
abelian result \eqn{quarticfermions} generalises to the non-abelian case.  
We want to go a  
little further and show that this is not only sufficient but actually  
necessary for the desired rearrangement to hold. We begin with the  
Fierz identity \eqn{fierztwo} with $\c\to\c^a$, $\psi\to D_\n\c^b$,  
$\l\to D_\m \c^c$ and $\varphi\to \c^d$: 
\ba\label{fierzbegin} 
\cb^a\g_{(\m}D^\n\c^b \ D^\m \cb^c\g_{\n)}\c^d 
&=&  
-{1\over 8} \cb^a\g_{(\m}\c^d \ D^\m \cb^c\g_{\n)}D^\n\c^b 
+{1\over 16}  \cb^a\g_\n\c^d D_\m\cb^c\g^\n D^\m\c^b 
\cr 
&&+{1\over 16}  \cb^a\g^{\r\s}_{\phantom{\m\n} (\m}\c^d  
\ D^\m \cb^c\g_{\n)\r\s}D^\n\c^b  
-{1\over 96} \cb^a\g^{\r\s\l}\c^d D_\m\cb^c \g_{\r\s\l} D^\m\c^b 
\cr 
&&-{1\over 384}  \cb^a\g^{\r\s\l\k}_{\phantom{\r\s\l\k} (\m}\c^d  
\ D^\m \cb^c\g_{\n)\r\s\l\k}D^\n\c^b \cr 
&&+{1\over 3840}  \cb^a\g^{\m\r\s\l\k}\c^d \ D_\n \cb^c\g_{\m\r\s\l\k}D^\n\c^b  
\ . 
\ea 
The first term on the r.h.s. vanishes on-shell as does the l.h.s. when $\m$  
and $\n$ are exchanged. The other terms can be simplified  
using the on-shell condition and partial integration so that 
\ba\label{fierznext} 
\cb^a\g_\m D_\n\c^b \ D^\m \cb^c\g^\n\c^d 
&\simeq& 
{1\over 2} \left( \cb^{(a} \g_\r D_\s \c^{d)}  
- \cb^{(a}\g_\s D_\r \c^{d)}\right) 
\cb^c \g^\s D^\r \c^b 
-{1\over 4} \cb^{[a} \g_\s D_\r \c^{d]}\ \cb^c \g^\s D^\r \c^b\cr 
&& 
+{1\over 24} \cb^{a} \g_{\r\s\l} D_\k \c^{d} \cb^c \g^{\r\s\l} D^\k \c^b  
-{1\over 8} \cb^{[a} \g_{\r\s\n} D_\m \c^{d]}\cb^c \g^{\r\s\m}D_\n\c^b \cr 
&& 
-{1\over 960}\cb^{[a}\g_{\m\r\s\l\k}D_\n\c^{d]}\ \cb^c\g^{\m\r\s\l\k}D^\n\c^b  
\ . 
\ea 
While the first and second terms exhibits the desired form and the  
third term can be  
dealt with by using again the Fierz identity \eqn{fierzone}, the fourth term  
is as troublesome (if not more) as the initial  
$\cb^a\g_\m D_\n\c^b \ D^\m \cb^c\g^\n\c^d$ we want to get rid of. However  
it comes antisymmetrised in $a$ and $d$, as does the fifth term, 
so that the symmetric part of eq. \eqn{fierznext} simply  
reduces to, using \eqn{fierzone} again, 
\be\label{fierzfinal} 
\cb^a\g_\m D_\n\c^b \  \cb^c\g^\n D^\m\c^d + ( a\leftrightarrow d) 
\simeq {2\over 3}\ \cb^a\g_\m D_\n\c^b\  \cb^c \g^\m D^\n\c^d 
+ ( a\leftrightarrow d) \ . 
\ee 
Thus we will get rid of the troublesome term and be able to use   
eq. \eqn{fierzfinal} 
provided $h_{abcd}$ is symmetric under exchange of $a$ and $d$ which we  
assume from now on. But as noted above, this implies that $h_{abcd}$  
is completely symmetric in all its indices: 
\be\label{hident} 
h_{abcd} = h \str \l_a\l_b\l_c\l_d 
\ee 
Hence, the contribution to the amplitude of the second term in  
${\cal L}_{\rm 4f}$ is 
\be\label{a44fh} 
A_4^{\rm 4f}\vert_{h-{\rm terms}} 
=-4 i\at^2 h\, \str\l_a\l_b\l_c\l_d\  
\left( s\ \ub_1 \g_\m u_4 \ub_2 \g^\m u_3  
- u\ \ub_1 \g_\m u_2 \ub_3 \g^\m u_4 \right) \ . 
\ee 
 
Matching the sum of both contributions \eqn{a44fgbis} and \eqn{a44fh} 
to the order $\ap^2$ four fermion string amplitude (recall that  
$\at=2\pi g\ap$) we get the condition 
\be\label{fourfermmatch} 
2g_1+g_3+2h = -{1\over 8} \ . 
\ee 
As a consistency check, we note that the {\it abelian} Born-Infeld  
action \eqn{bi} corresponds to $g_1=g_3={1\over 8}$ and $h=-{1\over 4}$  
which do satisfy this relation. With this in mind we parametrise 
\be\label{ghredef} 
g_1={1\over 8}\left(1 + \delta g +{4\over 3} \delta h \right) \quad , \quad 
g_3={1\over 8}\left(1 - 2 \delta g + {4\over 3} \delta h\right)\quad , \quad 
h=-{1\over 4}( 1+  \delta h) \ . 
\ee 
Similarly to what happened for the 2 boson / 2 fermion amplitude, matching of  
the 4 fermion amplitude does not completely determine the $U(N)$ tensor 
structure. 
Explicitly, we have found that the string 4 fermion amplitude is reproduced 
for any of the following interactions with arbitrary $\delta g$ and  
$\delta h$: 
\ba\label{fourfermexpl} 
{\cal L}_{\rm 4f} &=&  
 \str \left(  
{\at^2\over 8}\left(1+\delta g +{4\over 3}\delta h \right)\,  
\cb\g^\m D^\n\c\, \cb\g_\m D_\n \c 
-{\at^2\over 4}(1+\delta h )\, \cb\g^\m D^\n\c\, \cb\g_\n D_\m \c 
\right) \cr 
&&- {\at^2\over 32} \delta g\  d_{ade}d_{bce} 
\cb^a \g^\m D^\n \c^b\, \cb^c \g_\m D_\n \c^d \ . 
\ea 
Note that the parameter $\delta h$ does not reflect a lack of knowledge of  
the precise form of the action, but it only expresses  
the freedom to use the ``Fierz" identity \eqn{fierzfinal} to write the same  
term in two different ways: we may choose any $\delta h$ and still have the  
same action. The free parameter $\delta g$ corresponds to a contribution
that vanishes on-shell. It is proportional to (using (\ref{jacobi}))
\begin{equation}
  (f_{ace}f_{bde}+f_{abe}f_{cde})
     \cb^a \g^\m D^\n \c^b\, \cb^c \g_\m D_\n \c^d \,. 
\end{equation}
The first term we rewrite, using a Fierz transformation and contracting
 $\g$-matrices, in the form
\be
  -\half f_{ace}f_{bde} \cb^c \g^\m\c^a  D^\n\cb^b\g_\m D_\n\c^d\,.
\ee
In the second term we do a partial integration, obtaining (up to
terms that vanish on-shell)
\be
  -\half f_{abe}f_{cde} \cb^a\g^\m \c^b D^\n\cb^c\g_\m D_\n\c^d\,.
\ee
The two expressions are now in the same form, and can be seen to cancel
after renaming the indices. Therefore also the four-fermion terms
in the effective action are determined (up to contributions that
vanish on-shell) by the corresponding string amplitude.

\subsection{The string effective action } 
 
Finally we are in a position to collect our results and give the effective  
action up to and including all order $\ap^2$ terms, bosonic, fermionic and  
mixed.\footnote{ 
As already emphasized, we have nothing to say about a possible term  
$\sim \ap^2 F \cb\g\c\, \cb\g\c$ which would only show up in a  
five-point amplitude.} 
Without loss of generality we choose $\delta h=0$. Then  
the effective action reads: 
\ba\label{fulleffaction} 
{\cal L}_{\rm string} &=& \str \Bigg(  
- {1\over 4} F_{\m\n}F^{\m\n} +{i\over 2} \cb \g^\m D_\m \c 
+ {\at^2\over 8} F_{\m\n}F^{\n\r}F_{\r\s}F^{\s\m}  
- {\at^2\over 32} \left( F_{\m\n}F^{\m\n}\right)^2\cr 
&&\phantom{\str\Bigg(}  
+i{\at^2\over 4} \cb \g_\m D_\n \c F^{\m\r} F_\r^{\ \n}  
-i{\at^2\over 8} \cb \g_{\m\n\r} D_\s \c F^{\m\n} F^{\r\s}\cr 
&&\phantom{\str\Bigg(}  
+ {\at^2\over 8} \cb\g^\m D^\n\c\, \cb\g_\m D_\n \c 
- {\at^2\over 4} \cb\g^\m D^\n\c\, \cb\g_\n D_\m \c \Bigg)  
+ {\cal O}(\ap^3 g^2, \ap^2 g^3) \ . 
\ea 
Obviously, in the abelian limit this reduces to the standard abelian  
Born-Infeld action \eqn{biafter} after the field redefinition  
\eqn{abelfieldredef}. But the comparison with the Born-Infeld action as  
obtained by expanding the determinant goes further. Indeed, this non-abelian 
string effective action coincides with the result of the following  
manipulation: 
Take the abelian Born-Infeld action and expand it up to and including  
order $\ap^2$. Make the field redefinition to eliminate the order $\ap$  
term, and drop all ``on-shell" terms $\sim \at^2 \ds\c$. This gives  
\eqn{biafter}. Only then proceed to the obvious non-abelian generalisation  
and take a symmetrised trace. As noted above, this is not the same as  
taking the symmetrised trace before the field redefinition. 
This correct procedure might be called the modified symmetrised  
trace prescription. 
Note that it is unlikely that some sort of modified symmetrised trace  
prescription continues to hold at higher orders in $\ap$.

At this point it is useful to compare with the results of 
\cite{goteborg}. There, the $d=10$  super Yang-Mills 
 action through order $\ap^2$ 
was also determined by requiring linear supersymmetry. The claim is  
that the result is essentially unique. 
While the Lorentz structure is completely fixed there remains some  
small freedom in the adjoint structure, but again the only choice consistent 
with string theory turns out to be a symmetrised trace. 
If this uniqueness claim is correct, the action given in ref. 
 \cite{goteborg} and our string effective action (\ref{fulleffaction}) 
must coincide (up to on-shell terms and total derivatives). 
As we will now show, this is indeed the case. 
When using the same normalisation as ours, the action of 
ref.~\cite{goteborg} becomes   
\ba\label{fullsusyeffaction} 
{\cal L}_{\rm susy} &=& \str \Bigg(  
- {1\over 4} F_{\m\n}F^{\m\n} +{i\over 2} \cb \g^\m D_\m \c 
+ {\at^2\over 8} F_{\m\n}F^{\n\r}F_{\r\s}F^{\s\m}  
- {\at^2\over 32} \left( F_{\m\n}F^{\m\n}\right)^2\cr 
&&\phantom{\str\Bigg(}  
+i{\at^2\over 4} \cb \g_\m D_\n \c F^{\m\r} F_\r^{\ \n}  
-i{\at^2\over 8} \cb \g_{\m\n\r} D_\s \c F^{\m\n} F^{\r\s}\cr 
&&\phantom{\str\Bigg(}  
-{1\over 1440}\, {\at^2}  
\cb\g^{\m\n\rho}\chi D_\sigma\cb\, \g_{\m\n\rho} D^\s \c 
-{3\over 80}\, {\at^2} \cb\g^{\m\n\rho}\c \, D_\m\cb\g_\n D_\rho \c \cr 
&&\phantom{\str\Bigg(} 
-{7\over 480}{\at^2} g \, F^{\m\n}\cb\g_{\m\n\rho}\c\{\cb,\g^\rho\c\} 
+{1\over 2880}{\at^2} g \, F^{\m\n}\cb \g^{\rho\sigma\tau}\c 
\{\cb,\g_{\m\n\rho\sigma\tau} \c\} \Bigg) 
+ {\cal O}(\at^3) \ ,\cr 
&& 
\ea 
The first two lines of eqs.~(\ref{fullsusyeffaction}) and  
eq. (\ref{fulleffaction}) agree. It is clear that a direct (string) 
calculation of the 
last line of eq. (\ref{fullsusyeffaction}) which is of order $\ap^2 g^3$
would require the calculation of 
five point scattering amplitudes, so we have nothing to say about 
it here. What remains is the third line of  (\ref{fullsusyeffaction}) 
which has to be compared with the last line in  (\ref{fulleffaction}). 
Using the following identities (cf. \eqn{fierzfinal}  
and \eqn{fierzone}) 
\ba\label{osrels} 
\str \cb\g^\m D^\n\c\, \cb\g_\n D_\m \c 
&\simeq& {2\over 3}\, \str  \cb\g^\m D^\n\c\, \cb\g_\m D_\n \c \cr 
\str \cb\g^{\m\n\rho}\chi D_\sigma\cb\, \g_{\m\n\rho} D^\s \c 
&=&24\ \str  \cb\g^\m D^\n\c\, \cb\g_\m D_\n \c  \cr 
\str \cb\g^{\m\n\rho}\c \, D_\m\cb\g_\n D_\rho \c  
&\simeq&{2\over 3}\, \str \cb\g^\m D^\n\c\, \cb\g_\m D_\n \c \ , 
\ea 
we find that the third line of 
eq. (\ref{fullsusyeffaction}) 
agrees with  
the third line of 
eq. (\ref{fulleffaction}), provided 
\be\label{gotematch} 
-{1\over 1440} \times 24 
-{3\over 80} \times {2\over 3} 
= {1\over 8}  -{1\over 4}\times {2\over 3}   \ , 
\ee 
which indeed is true.

In \cite{goteborg}, the presence of a non-linear 
supersymmetry of the action (\ref{fullsusyeffaction}) 
was established as well. This provided strong 
consistency checks on various terms  
although the values of the coefficients of  
the four-fermion terms 
are insensitive to this. Indeed, one easily checks that the variation 
of these two terms under the non-linear supersymmetry result in expressions 
proportional to equations of motion. Happily, as just checked, these terms  
are precisely equivalent to the four-fermion terms in the string effective  
action, which now provides an independent check.

\section{Kappa-symmetry} 
 
The purpose of this section is to compare the results for the 
effective action obtained in 
Section 2 with the results that follow from the  
requirement of $\kappa$-symmetry. In the abelian case, $\kappa$-symmetry 
has led, in the limit of constant $F$, to exact answers for D-branes 
in a flat \cite{Aga} as well as a curved \cite{curved} background.

Another reason to reconsider the results obtained in \cite{BdRS}, 
is the recent claim  \cite{goteborg}, 
that supersymmetric Yang-Mills theory in $d=10$, to order $\alpha'^2$,  
must contain a symmetric trace of the 
Yang-Mills generators. According to \cite{goteborg}  
any deviation from the symmetric trace must 
be trivial, in the sense that it can be removed by a field 
redefinition. 
 
The results of \cite{BdRS} indicate  
that, as far as the terms bilinear in the fermions are concerned, 
a nontrivial deviation from the symmetric trace 
does occur. Since quartic fermions were not considered in  
\cite{BdRS}, we will disregard them in this section. 
The results of \cite{BdRS} are based on the assumption that 
a particular non-abelian version of $\kappa$-symmetry exists. 
This non-abelian $\kappa$-symmetry automatically leads to linear 
and nonlinear supersymmetries after $\kappa$-gauge fixing. The 
non-abelian $\kappa$-symmetry proposal of \cite{BdRS} was only established 
at order $F^2$ in the variation.  
This implies that after $\kappa$-gauge fixing the linear 
supersymmetry has only been established for the  $F^2$ terms in the 
action  but not for the $\alpha'^2F^4$ terms (there are no $F^3$ terms). 
The action of  
\cite{BdRS} also contains terms which are of the form  
$\alpha'^2\bar\theta\partial\theta F^2$. These terms are needed to realize 
the nonlinear supersymmetry at order $F^2$. 
On the other hand, the linear supersymmetry calculation of \cite{goteborg} 
was performed up to order $\alpha'^2$ in the variation. This fixes the 
$\alpha'^2F^4$ terms in the action which, by linear supersymmetry, 
are connected to the $\alpha'^2\bar\theta\partial\theta F^2$ terms. 
 
The apparent contradiction between \cite{goteborg} and \cite{BdRS}  is that 
the linear supersymmetry calculation 
 of \cite{goteborg} leads to a symmetric trace 
prescription of the  $\alpha'^2\bar\theta\partial\theta F^2$ terms 
in the effective action whereas the $\kappa$-symmetry  
calculation of \cite{BdRS} 
shows that these terms do not satisfy the symmetric trace prescription.

We should keep in mind that since $\kappa$-symmetry has only been established 
up to order $F^2$ terms in the variation, we have no 
guarantee that we can proceed to higher orders. Indeed, the results of 
\cite{goteborg} indicate that proceeding with the $\kappa$-symmetry  
calculation to the next order might be problematic. 
 
Strictly speaking there are two possible situations: 
 
\begin{description} 
\item{(1)} It is possible that after redefinitions the  
      $\alpha'^2\bar\theta\partial\theta F^2$ terms of \cite{BdRS} 
       do become a symmetric trace, in which case the result agrees with 
       \cite{goteborg}.  
\item{(2)} If it is not a symmetric trace, under any field-redefinition, 
      then, assuming that the conclusion of \cite{goteborg} is correct,  
      $\kappa$-symmetry must fail at the next order. 
\end{description} 
 
We will show in the remainder of this section that the first possibility 
does not apply. There are no field redefinitions under which all terms 
in the action of \cite{BdRS} can be written as a symmetrised trace. 
We are left with the second possibility and, indeed, we will show that  
$\kappa$-symmetry fails at order $F^3$ in the 
variation. The consequences of this will be discussed in the next section.

\subsection{kappa-invariant action}

It is convenient to first reformulate the results of  
\cite{BdRS} in the form obtained after 
making the field redefinitions discussed in Section 2. For the 
$\kappa$-symmetric formulation these redefinitions take the form: 
\begin{eqnarray} 
 && \bar\theta^a\to \bar\theta^{\prime\,a} -  
   \noverm{1}{8}\at d^{abc} \,\bar\theta^{\prime\,b}\sigma_3\gamma\cdot F^c P_-\,, 
 \\ 
 && A_\mu^a\to A_\mu^{\prime\,a} -  \noverm{i}{8} \at
  d^{abc} \,\bar\theta^{\prime\,b} 
          \gamma_{11}(i\sigma_2)\gamma_\mu\theta^{\prime\,c}\,. 
\end{eqnarray} 
Here we use the following notation:
\begin{equation} 
  P_\pm = \half (1\pm \gamma_{11}\sigma_1)\,. 
\end{equation} 
These are projection operators, and satisfy 
\begin{equation} 
  P_\pm P_\pm=P_\pm\,,\ P_-P_+=0\,,\  
  P_-\sigma_3 = \sigma_3P_+\, . 
\end{equation}

After this redefinition the action is 
\begin{eqnarray} 
  {\cal L} &=&  i\bar\theta^a P_- \gamma^\mu\Dpartial_\mu\theta^a 
               -\quart F_{\mu\nu}^aF^{\mu\nu\,a} 
\nonumber\\ 
&& 
  +\noverm{i}{8} \at^2 d^{ae(c}d^{d)be} \bar\theta^aP_-\gamma_\mu\Dpartial_\nu\theta^b 
      T^{\mu\nu\,cd} 
\nonumber\\ 
&& - \noverm{i}{16}\at^2   d^{ae[c}d^{d]be} \bar\theta^a P_- 
                      \gamma_{\mu\nu\rho} 
      \{\Dpartial^\rho\theta^b F^{\mu\sigma\,c} F_\sigma{}^{\nu\,d} 
         - \Dpartial_\sigma\theta^b F^{\mu\nu\,c}F^{\rho\sigma\,d}\} 
\nonumber\\ 
\label{Lquad} 
&&    
   +\noverm{i}{64}\at^2 
   d^{ace}d^{bde}\bar\theta^aP_-\gamma_{\mu\nu\rho\sigma\tau} 
   \Dpartial_\tau\theta^b F^{\mu\nu\,c}F^{\rho\sigma\,d}\,, 
\end{eqnarray} 
where $T^{\mu\nu\,cd}$ is the nonabelian generalization of the  
energy-momentum tensor: 
\begin{equation} 
  T^{\mu\nu\,cd} =  F^{\mu\rho\,(c}F_{\rho}{}^{d)\nu} +  
         \quart \eta^{\mu\nu} F_{\rho\sigma}^cF^{\rho\sigma\,d}\,. 
\end{equation} 
It is invariant under the following $\kappa$-symmetry transformations: 
\begin{eqnarray} 
\label{deltath} 
  \delta\bar\theta^a &=& \bar\eta^a - \bar\epsilon^a 
   +\noverm{1}{8} \at d^{abc}\,(\bar\eta^b - \bar\epsilon^b) 
     \sigma_3\gamma\cdot F^c P_-\,, 
 \\ 
   \delta A_\mu^a &=&  \noverm{i}{2}\at d^{abc}\,\bar\eta^b  
P_-\sigma_3\gamma_\mu\theta^c  
    + \noverm{i}{2}\at d^{abc}\,\bar\epsilon^b P_+ \sigma_3\gamma_\mu\theta^c 
\nonumber\\ 
&& +\noverm{i}{8}\at^2 d^{ae(c}d^{d)be} \bar\epsilon^b P_- 
       \gamma_\rho\theta^c F^\rho{}_\mu{}^d 
\nonumber\\ 
&& +\noverm{i}{16}\at^2  d^{ae[c}d^{d]be}\bar\epsilon^b P_- 
       \gamma_{\mu\nu\rho}\theta^c F^{\nu\rho\,d} 
\nonumber\\ 
&& +\noverm{i}{32}\at^2 (3d^{ace}d^{bde}+d^{abe}d^{cde})\,  
    \bar\eta^b(P_++P_-)\gamma_k\theta^cF^k{}_i{}^d 
\nonumber\\ 
&& +\noverm{i}{16}\at^2  d^{ae[c}d^{d]be}\bar\eta^b(P_+-P_-)\gamma_\rho\theta^c 
F^\rho{}_\mu{}^d 
\nonumber\\ 
&& +\noverm{i}{16}\at^2  d^{ace}d^{bde} \bar\eta^bP_+\gamma_{\mu\nu\rho}\theta^c 
F^{\nu\rho\,d} 
\nonumber\\ 
\label{deltaV} 
&& + \noverm{i}{16}\at^2  d^{ae[c}d^{d]be} \bar\eta^bP_-\gamma_{\mu\nu\rho}\theta^c 
F^{\nu\rho\,d}\,, 
\end{eqnarray} 
where the parameter $\eta^a$ is of the form 
\begin{equation} 
  \bar\eta^a = \bar\kappa^b( \delta^{ab} + \Gamma^{ab})\,. 
\end{equation} 
The matrix $\Gamma$ must square to one, and can be reconstructed  
in the present basis from the results given in \cite{BdRS}. The parameter 
$\epsilon^a$ is constant, and must satisfy $f^{abc}\epsilon^c=0$. 
 
\subsection{Gauge fixing and supersymmetry} 
 
Gauge-fixing follows the same lines as  
discussed in \cite{BdRS}. The $\kappa$-symmetry 
is gauge-fixed by setting $\theta_2^A=0$, and the remaining 
symmetries are linear and nonlinear supersymmetry.  
We will 
present only the results. After gauge-fixing the action reads: 
\begin{eqnarray} 
  {\cal L} &=&  \ihalf \bar\chi^a \gamma^\mu\Dpartial_\mu\chi^a 
               -\quart F_{\mu\nu}^aF^{\mu\nu\,a} 
\nonumber\\ 
&& 
  +\noverm{i}{16}\at^2  d^{ae(c}d^{d)be} \bar\chi^a \gamma_\mu\Dpartial_\nu\chi^b 
      T^{\mu\nu\,cd} 
\nonumber\\ 
&& - \noverm{i}{32}\at^2    d^{ae[c}d^{d]be} \bar\chi^a  
                      \gamma_{\mu\nu\rho} 
      \{\Dpartial^\rho\chi^b F^{\mu\sigma\,c} F_\sigma{}^{\nu\,d} 
         - \Dpartial_\sigma\chi^b F^{\mu\nu\,c}F^{\rho\sigma\,d}\} 
\nonumber\\ 
\label{Lgfquad} 
&&    
   +\noverm{i}{128}\at^2 d^{ace}d^{bde}\bar\chi^a \gamma_{\mu\nu\rho\sigma\tau} 
   \Dpartial_\tau\chi^b F^{\mu\nu\,c}F^{\rho\sigma\,d}\,. 
\end{eqnarray} 
The transformation rules under supersymmetry simplify because 
of the condition $f^{abc}\epsilon^c=0$. This means we can 
choose a basis in the $U(N)$ Lie-algebra such that only 
one $\epsilon$, corresponding to the $U(1)$ direction, 
remains. Setting $a=0$ for the $U(1)$ direction, we then 
use $d^{ab0} = \delta^{ab}$ (up to a  constant, which we absorb into the normalisation of $\epsilon$).  
The transformation rules then take on the following form: 
\begin{eqnarray} 
\label{delchi} 
   \delta\bar\chi^a &=& - (\bar\epsilon_1+\bar\epsilon_2)\delta^{a0} 
      -  \noverm{1}{8}\at (\bar\epsilon_1-\bar\epsilon_2)\gamma\cdot F^a\,, 
\\ 
\label{delV} 
   \delta A_\mu^a &=& +\iquart \at (\bar\epsilon_1-\bar\epsilon_2) \gamma_\mu\chi^a 
   +\noverm{i}{8}\at^2 d^{acd} (\bar\epsilon_1+\bar\epsilon_2) 
        \gamma_\rho\chi^cF^\rho{}_\mu{}^d\,. 
\end{eqnarray} 
 
The algebra of the linear supersymmetry is as usual. The nonlinear 
supersymmetry gives a covariant translation on $A$ 
(the same as for linear supersymmetry). To see this on 
$\chi$ would require the presence of higher-order fermions in the  
transformation rule of $\chi$, but these have not been determined. 
The algebra of linear with nonlinear supersymmetry gives a 
constant shift on the $U(1)$ vector. On $\chi$ this  
commutator also requires higher-order fermion contributions. 
 
The action (\ref{Lgfquad})  
can be simplified somewhat by redefining $\chi^a$ 
with $F^2$-dependent terms. This gives: 
\be\label{Lgfquad2} 
{\cal L} = \ihalf \bar\chi^a \gamma^\mu\Dpartial_\mu\chi^a 
               -\quart F_{\mu\nu}^aF^{\mu\nu\,a} 
+\noverm{i}{16}\at^2 d^{ace}d^{bde} \bar\chi^a \gamma_\mu\Dpartial_\nu\chi^b 
        F^{\mu\rho\,c}F_\rho{}^{\nu\,d} 
 - \noverm{i}{32}\at^2  d^{ade}d^{bce} \bar\chi^a  
                      \gamma_{\mu\nu\rho} 
         \Dpartial_\sigma\chi^b F^{\mu\nu\,c}F^{\rho\sigma\,d}\, . 
\ee
In this form the result can be most easily compared with the results 
of Section 2. 
Note that (\ref{Lgfquad2}) is not a symmetric trace and therefore 
it differs from the action (\ref{fulleffaction}) we found in Section 2. 
This is an 
aspect of the $\kappa$-symmetric formulation which now is seen to be 
independent of the redefinition we performed. Of course, in the 
abelian limit ($d^{abc} \rightarrow 2$), the action 
(\ref{Lgfquad2}) should coincide with the action (\ref{fulleffaction})  
of Section 2, as it does. 
 
At order $\alpha'^2$ we can only check the nonlinear supersymmetry, and it is 
indeed valid. For the linear supersymmetry at order $\alpha'^2$ we 
need also the $F^4$ term. In fact, long ago,  
in \cite{BRS}, it was shown that the  
following action (ignoring quartic fermions), 
which has a symmetric trace, is invariant under 
linear supersymmetry: 
\begin{eqnarray} 
    {\cal L}  &=&  
  \noverm{1}{12}(d^{abe}d^{cde}+d^{ade}d^{bce}+d^{ace}d^{bde}) \at^2\times 
\nonumber\\ 
&& \times\   \big(\noverm{1}{8}F_{\mu\nu}^aF^{\nu\rho\,b}F_{\rho\sigma}^c 
F^{\sigma\mu\,d} 
                -\noverm{1}{32} F_{\mu\nu}^aF^{\mu\nu\,b}F_{\rho\sigma}^c 
F^{\rho\sigma\,d} 
\nonumber\\ 
\label{LquadStr} 
&&\quad +\iquart  \bar\chi^a \gamma_\mu\Dpartial_\nu\chi^b 
      T^{\mu\nu\,cd}  - \noverm{i}{8}   \bar\chi^a  
                      \gamma_{\mu\nu\rho} 
      \{\Dpartial^\rho\chi^b F^{\mu\sigma\,c} F_\sigma{}^{\nu\,d} 
         - \Dpartial_\sigma\chi^b F^{\mu\nu\,c}F^{\rho\sigma\,d}\} 
\nonumber\\ 
&&\quad   +\noverm{i}{32} \bar\chi^a \gamma_{\mu\nu\rho\sigma\tau} 
   \Dpartial_\tau\chi^b F^{\mu\nu\,c}F^{\rho\sigma\,d}\big)\,. 
\end{eqnarray} 
Performing an analogous redefinition of $\c^a$ with $F^2$-dependent terms, as
above, this can be rewritten as
\ba\label{LquadStrtwo} 
 {\cal L}  &=&  
  \noverm{1}{12}(d^{abe}d^{cde}+d^{ade}d^{bce}+d^{ace}d^{bde}) \at^2\times 
\cr
&& \times\   \big(
\noverm{1}{8}F_{\mu\nu}^aF^{\nu\rho\,b}F_{\rho\sigma}^c F^{\sigma\mu\,d} 
-\noverm{1}{32} F_{\mu\nu}^aF^{\mu\nu\,b}F_{\rho\sigma}^c F^{\rho\sigma\,d} 
\cr
&&\quad +\iquart  \bar\chi^a \gamma_\mu\Dpartial_\nu\chi^b 
      T^{\mu\nu\,cd}  - \noverm{i}{8}   \bar\chi^a  
                      \gamma_{\mu\nu\rho} 
     \Dpartial_\sigma\chi^b F^{\mu\nu\,c}F^{\rho\sigma\,d}\big) \, . 
\ea
 
We now want to check that our action (\ref{Lgfquad2})  
can be made invariant under linear supersymmetry 
after adding the well-known STr $F^4$ terms predicted by  
string theory. 
The simplest way to check this is to add to  
(\ref{Lgfquad2}) a symmetric trace $F^4$ term, with the correct 
normalization, and then to subtract the result from 
(\ref{LquadStrtwo}). This difference, 
${\cal L}_{\rm rest}$, should then also be supersymmetric. 
This can only happen if the variation of this difference can be 
cancelled by new order $\alpha'^2$-variations of the fields $\chi$ and 
$A$. This requires that all terms in the variation can be rewritten in 
terms of the (lowest order) equations of motion of these fields. For  
this analysis it is of course crucial that there are no other parts 
of the action which could interfere with this calculation, such as 
higher-derivative terms. We have verified that to this order  
higher-derivative contributions can always be reexpressed in terms of  
lowest-order equations of motions, and can be eliminated by field  
redefinitions. 
 
The variation of ${\cal L}_{\rm rest}$ 
under linear supersymmetry, in which case  
$\epsilon\equiv \epsilon_1-\epsilon_2$, is 
\begin{eqnarray} 
 \delta{\cal L}_{\rm rest} &=&  
  \at^3\,(\noverm{i}{16} 
  \bar\epsilon\gamma_{\rho\sigma}\gamma_\mu\Dpartial_\nu \chi^a 
       F^{\rho\sigma\,b} F^\mu{}_\tau{}^c F^{\tau\nu\,d}\, (P-Q)^{abcd} 
\nonumber\\ 
\label{d3Lrest} 
&& +\noverm{i}{32} \bar\epsilon\gamma_{\rho\sigma} \gamma_{\mu\nu\tau} 
        \Dpartial_\lambda\chi^a F^{\rho\sigma\,b}  
F^{\mu\nu\,c}F^{\tau\lambda\,d}  
\,(P+Q)^{abcd} )\,. 
\end{eqnarray} 
where the tensors $P$ and $Q$ have been defined in the Appendix. To analyze 
this variation, it is convenient to multiply all $\gamma$-matrices  
together in terms of a $\gamma^{(5)}$, a $\gamma^{(3)}$, and a  
$\gamma^{(1)}$. Using the symmetry properties of $P$ and $Q$ it 
is not very complicated to show that the $\gamma^{(5)}$ contribution 
can be written in terms of equations of motion. However, this analysis 
fails at the level of the $\gamma^{(3)}$ terms. We found that for 
certain dimensions lower than ten (in particular $d=3$) 
the $\gamma^{(3)}$-terms can also be rewritten in terms of equations of  
motion, but in the general case, and in particular in $d=10$, this does 
not work. For $d=3$ the $\gamma^{(1)}$-terms still give problems, 
which can however be resolved by adding
$F^4$-terms which are not a symmetric trace. 
For $d=10$ we conclude that  
$\kappa$-symmetry fails at this order.

\section{Conclusions} 
 
In this paper we have determined the string effective action from the  
four-point string scattering amplitudes, including all fermionic terms through  
order $\ap^2 g^2$. We have also refined the determination of the  
$\k$-symmetric action of \cite{BdRS} by proceeding in a way which  
yields no order $\ap$ term from the beginning, so this corresponds 
to the situation after the field redefinition. The two results do not coincide. 
While $\k$-symmetry might be desirable, it is not a sacred principle.  
On the other hand, the effective action (\ref{fulleffaction}) 
we obtained by matching string  
amplitudes really is the true string effective action. As  
repeatedly mentioned, its order $\ap^2$ terms are only determined up  
to on-shell terms, but this is precisely the freedom we have to perform  
further field redefinitions of order $\ap^2$. 
In ref. \cite{goteborg} a super Yang-Mills action through order $\ap^2$ 
including all fermionic terms  
was also determined recently by requiring linear supersymmetry. The claim  
of \cite{goteborg} is  
that the result is essentially unique and we have shown that it 
coincides with the string effective action we have determined.  
For completenes we give here the result where we have rewritten 
the quartic fermions as a single term, using the identities 
(\ref{osrels}): 
\ba\label{finalaction} 
{\cal L}_{\rm string} &=& \str \Bigg(  
- {1\over 4} F_{\m\n}F^{\m\n} +{i\over 2} \cb \g^\m D_\m \c 
+ {\at^2\over 8} F_{\m\n}F^{\n\r}F_{\r\s}F^{\s\m}  
- {\at^2\over 32} \left( F_{\m\n}F^{\m\n}\right)^2\cr 
&&\phantom{\str\Bigg(}  
+i{\at^2\over 4} \cb \g_\m D_\n \c F^{\m\r} F_\r^{\ \n}  
-i{\at^2\over 8} \cb \g_{\m\n\r} D_\s \c F^{\m\n} F^{\r\s}\cr 
&&\phantom{\str\Bigg(}  
-{\at^2\over 24} \cb\g^\m D^\n\c\, \cb\g_\m D_\n \c \Bigg) 
+ {\cal O}(\ap^3 g^2, \ap^2 g^3) \ .\cr 
&& 
\ea 
We have omitted the $F\chi^4$ term since it is $\O(\ap^2 g^3)$ and 
would only show up in the 
calculation of the five-point amplitude.

We can safely conclude that the symmetrised 
trace prescription for the non-abelian Born-Infeld action holds  
through order $\ap^2$, 
including {\em all} fermionic and derivative terms. As we pointed out in 
section 2.4, one should be careful with the field redefinitions. The 
redefinitions should be done before implementing the symmetrised 
trace! We also stress that the present conclusion does not imply that 
the symmetrised trace prescription will continue to hold at higher 
orders. In fact a closer investigation of the $\ap$-expansion of the string  
scattering amplitudes \cite{ab2} indicates that the 
symmetrised trace prescription will fail beyond order $\ap^2$.

Finally, we found that $\kappa$-symmetry cannot be extended to 
the order $F^3$ in the variation.  
On the other hand, the fact that the 
effective action through order $\ap^2$ shows both a linear and a 
non-linear supersymmetry is indicative for the existence of an 
underlying $\kappa$-invariant formulation.  
The work of \cite{BdRS} was based on a 
non-abelian $\kappa$-symmetry, under which all fermions transform, such 
that the $\kappa$-parameter is also in the adjoint representation of the
Yang-Mills group. It may be that this approach has been too ambitious, 
and that 
only a single $\kappa$-symmetry can be realised. It is also conceivable that
the approach of \cite{BdRS} was not ambitious enough and, maybe, besides 
nonabelian $\kappa$-transformations, it is also required to introduce  
some kind of non-abelian diffeomorphisms on the worldvolume. 
Clearly more thought is required before $\kappa$-symmetry, in this context, 
is finally put to rest.

\vskip 1.cm 
\centerline{\bf Acknowledgements} 
 
We are grateful for fruitful discussions with M. Cederwall, L. De Foss\'e,  
J.-P. Derendinger, P.~Koerber,  
R. Russo, K. Sfetsos and J. Troost. We thank the Institute of Physics at  
Neuch\^atel and the Institute for Theoretical Physics at Groningen  
for hospitality. This work is supported by the European Commission  
RTN programme  
HPRN-CT-2000-00131 in which E.B. and M.d.R. are associated with  
the University of Utrecht and A.S. is associated with the University  
of Leuven. 
\appendix 
 
\section{Conventions and useful identities} 
 
In this appendix we gather some conventions and identities we use. 
 
\underline{Kinematics:} 
\be s=(k_1+k_2)^2 \quad , \quad t=(k_1+k_3)^2 \quad , \quad u=(k_1+k_4)^2 
\ee 
with all momenta incoming and we use signature $(+,-,\ldots ,-)$. Since  
all our states are massless we have $s+t+u=0$. 
 
\underline{Spinors:} 
The Clifford algebra is $\{\g^\m,\g^\n\}=2 \eta^{\m\n}$, i.e. $(\g^0)^2=+1$. 
Antisymmetric products of $\g$-matrices are defined with weight 1:  
$\g_{\m\n}=\ha (\g_\m\g_\n-\g_\n\g_\m)$ etc. Often used identities are 
\ba\label{gammaalg} 
\g_{\m\n\r} 
&=&\g_\m\g_\n\g_\r-\g_\m \eta_{\n\r}+\g_\n \eta_{\m\r} -\g_\r \eta_{\m\n}\cr 
\g_\m\g_{\n\r}&=&\g_{\m\n\r}+\g_\r \eta_{\m\n} - \g_\n \eta_{\m\r}\cr 
\g_{\n\r}\g_\m&=&\g_{\n\r\m}+\g_\n \eta_{\m\r} -\g_\r \eta_{\m\n} 
\ea 
The ten-dimensional spinors are 16-component Majorana-Weyl spinors and  
satisfy various identities. In particular, due to the Weyl property 
$\cb_1 \g_{\m_1\ldots \m_p}\c_2=0$ for all even $p$, and the expressions  
with $p>5$ are related to those with $10-p<5$. Due to the Majorana  
property anticommuting spinor fields satisfy 
\ba\label{majident} 
 \cb_1 \c_2 &=& \cb_2 \c_1 \ , \quad 
\cb_1 \g_\m \c_2 = - \cb_2 \g_\m \c_1 \ , \cr 
\cb_1 \g_{\m_1 \ldots \m_p} \c_2  
&=& (-)^p \cb_2 \g_{\m_p \ldots \m_1} \c_1 = 
(-)^{p(p+1)/2}\cb_2 \g_{\m_1 \ldots \m_p} \c_1  
\ea 
Note that when the anticommuting spinor fields are replaced by commuting  
spinor wave-functions we have the analogous identities but with an extra  
minus sign. 
 
There are also various Fierz identities which can be derived from the  
following basic identity \cite{BRW} valid for ten-dimensional  Majorana-Weyl 
spinors (a Weyl projector is implicitly assumed to multiply the r.h.s.) 
\be\label{genfierz} 
\p\lb = -{1\over 16} \g^\m ( \lb\g_\m\p) 
+ {1\over 96}\g^{\m\n\r} (\lb \g_{\m\n\r}\p)  
- {1\over 3840}\g^{\m\n\r\s\k} (\lb \g_{\m\n\r\s\k}\p)  
\ee 
from which follows 
\be\label{fierzone} 
\cb\g^\m\p \ \lb\g_\m\vf = \ha \cb\g^\m\vf \ \lb\g_\m\p  
-{1\over 24}  \cb\g^{\m\n\r}\vf \ \lb\g_{\m\n\r}\p  
\ee 
as well as 
\ba\label{fierztwo} 
\cb\g^{(\m}\p \ \lb\g^{\n)}\vf &=&  
-{1\over 8} \cb\g^{(\m}\vf \ \lb\g^{\n)}\p 
+{1\over 16}  \cb\g^{\r\s (\m}\vf \ \lb\g^{\n)}_{\phantom{\n)}\r\s}\p  
-{1\over 384}\cb\g^{\r\s\l\k (\m}\vf \ \lb\g^{\n)}_{\phantom{\n)}\r\s\l\k}\p\cr 
&& + \eta^{\m\n} \left[ 
{1\over 16} \cb\g^\r\vf \ \lb\g_\r\p 
-{1\over 96} \cb\g^{\r\s\l}\vf \ \lb\g_{\r\s\l}\p 
+{1\over 3840} \cb\g^{\r\s\l\k\tau}\vf \ \lb\g_{\r\s\l\k\tau}\p \right] \ , 
\ea 
where $(\m  \n)$ indicates symmetrisation in $\m$ and $\n$.

\underline{Gauge group, $d_{abc}$ and $f_{abc}$ tensors :} 
We denote by $\l_a$ the hermitian generators of the fundamental  
representation of ${\rm U}(N)$. The various normalisations are fixed by  
\be\label{gaugegen} 
[\l_a,\l_b]=i f_{abc} \l_c \quad , \quad  
\{\l_a,\l_b\}= d_{abc} \l_c \quad , \quad  
\tr \l_a\l_b=\delta_{ab} 
\ee 
with real structure constants $f_{abc}$ and real $d_{abc}$. 
These definitions imply 
\be\label{threegentrace} 
\tr [\l_a,\l_b] \l_c = i f_{abc} \quad 
,\quad \tr \{\l_a,\l_b\}\l_c=d_{abc} \ . 
\ee 
The generators of the adjoint representation are  
$(T_a^{\rm adj})_{bc}=- i f_{abc}$, which is the only representation of  
interest to us. The covariant derivative then is 
\be\label{covder} 
(D_\m^{\rm adj})_{ac}= \delta_{ac}\d_\m - i g A_\m^b (T_b^{\rm adj})_{ac} 
=\delta_{ac}\d_\m + g f_{abc} A_\m^b 
\ee 
The field strength then is given by $[D_\m,D_\n]_{ac}= g f_{abc} F^b_{\m\n}$  
i.e. 
\be\label{fieldstr} 
F^a_{\m\n}=\d_\m A^a_\n - \d_\n A^a_\m + g f_{abc} A^b_\m A^c_\n 
\ee 
 
Possible 4-index tensors on the gauge group that could arise from a  
single trace 
are of the form $d_{abe}d_{cde}$, $f_{abe}f_{cde}$ or $d_{abe}f_{cde}$.  
There are 12 such possible tensors, but they are related by various 
Jacobi identities: 
\ba\label{jacobi} 
&&f_{abe}f_{cde}=d_{ace}d_{bde}-d_{ade}d_{bce} \cr 
&&d_{abe}f_{cde}+d_{bce}f_{ade}+d_{cae}f_{bde}=0 \ . 
\ea 
The first type of identities allows to express all $ff$ tensors as  
$dd$ tensors, and the second type of identities allows to express 3  
among the 6  $df$ tensors in terms of the 3 others. We may choose 
$\b_1=d_{abe}f_{cde}$, $ \b_2=d_{cde}f_{abe}$ and  
$\b_3=d_{ade}f_{bce}-d_{bde}f_{ace}$ as independent, and use them  
to express the three other $\b_4=d_{ace}f_{bde}$, $\b_5=d_{bce}f_{ade}$  
and $\b_6=d_{ade}f_{bce}+d_{bde} f_{ace}$: 
\be\label{dfidentities} 
\b_4=-(\b_1+\b_3)/2+\b_2\ ,\quad 
\b_5=-(\b_1-\b_3)/2-\b_2\ ,\quad 
\b_6=\b_1 \ . 
\ee 
Then, if we expand a general tensor as 
\be\label{gentensexp} 
X_{abcd}=x_1 d_{abe}d_{cde} + x_2 d_{ace}d_{bde} + x_3 d_{ade}d_{bce} 
+x_4 d_{abe}f_{cde}+x_5 d_{cde}f_{abe} + x_6(d_{ade}f_{bce}-d_{bde}f_{ace})\ , 
\ee 
knowing only $X_{(ab)cd}$ will leave $x_2-x_3$, $x_5$ and $x_6$  
undetermined, while knowing $X_{abcd}+X_{badc}$ will leave $x_4$  
and $x_5$ undetermined. 
Finally we note that 
\be\label{symtrdabc} 
\str\l_a\l_b\l_c\l_d= {1\over 12}  
\left(d_{abe}d_{cde} + d_{ace}d_{bde} + d_{ade}d_{bce} \right) \ . 
\ee 
 
In the text we have introduced the tensors $P$ and $Q$: 
\ba\label{pqtensors} 
P^{abcd} &\equiv&  
\noverm{1}{24}( d^{ace}d^{bde}+d^{ade}d^{bce}-2d^{abe}d^{cde}) 
= -\noverm{1}{24}(f^{ace}f^{bde}+f^{ade}f^{bce}) \cr 
Q^{abcd} &\equiv&  
\noverm{1}{8} (d^{ace}d^{bde}-d^{ade}d^{bce}) 
=\noverm{1}{8}\, f^{abe}f^{cde} 
\ea 
The combinations $P\pm Q$ are: 
\ba\label{pplusminusq} 
(P+Q)^{abcd} &=& \noverm{1}{12}(f^{abe}f^{cde}+f^{ade}f^{cbe}) \cr 
(P-Q)^{abcd} &=& -\noverm{1}{12}(f^{abe}f^{cde}+f^{ace}f^{bde})\ . 
\ea 
Note that $P+Q$ is symmetric in $bd$ and $ac$, $P-Q$ is 
symmetric in $bc$, $ad$.

\underline{Feynman rules:} From ${\cal L}_{\rm SYM} 
=\tr \left( -{1\over 4} F_{\m\n}F^{\m\n} +{i\over 2} \cb \g^\m D_\m \c\right)$  
we read the following Feynman rules for tree amplitudes (no ghosts):  
the fermion propagator is $+ i \delta_{ab}/ \ks$, the gluon propagator  
$-i \delta_{ab} \eta_{\m\n}/ k^2$ (any gauge dependent additional terms  
$\sim k_\m$ or $\sim k_\n$ drop out in all our amplitudes). All vertices  
are obtained from the relevant interaction terms with the rule  
$\d_\m \to - i k_\m$ where the momentum $k$ is going into the vertex.

\end{document}